\documentclass[12pt,preprint]{aastex}
\usepackage{graphicx,natbib,amsfonts,amssymb,lscape}
\usepackage[figuresright]{rotating}
\usepackage{apjfonts}
\usepackage{color}
\usepackage{lscape}     
\usepackage{graphicx}
\usepackage{graphics}
\usepackage{psfig}
\usepackage{multicol}
\usepackage{natbib}

\newcommand{\ha}{\ifmmode H\alpha \else H$\alpha$\ \fi}
\newcommand{\hb}{\ifmmode H\beta \else H$\beta$\ \fi}

\newcommand{\hbb}{\ifmmode H\beta^{b} \else H$\beta^{b}$\ \fi}
\newcommand{\hbn}{\ifmmode H\beta^{n} \else H$\beta^{n}$\ \fi}
\newcommand{\ergs}{\ifmmode {\rm erg\ s}^{-1} \else erg s$^{-1}$\ \fi}
\newcommand{\civ}{C {\sc iv}\ }
\newcommand{\ciii}{C {\sc iii}\ }
\newcommand{\mgii}{Mg {\sc ii}\ }

\newcommand{\kms}{\ifmmode {\rm km\ s}^{-1} \else km s$^{-1}$\ \fi}
\newcommand{\msun}{\ifmmode M_{\odot} \else $M_{\odot}$\ \fi}
\newcommand{\lv}{\ifmmode L_{\lambda}(5100\AA) \else $L_{\lambda}(5100\AA)$\ \fi}

\slugcomment{submitted to ApJ, on \today}
\usepackage{CJK}
\usepackage[T1]{fontenc}

\begin{document}
\begin{CJK}{UTF8}{gbsn}

\title{The optical variability of SDSS quasars from multi-epoch spectroscopy: I. Results from 60 quasars with $\geq$ six-epoch spectra}
\author{Hengxiao Guo (郭恒潇)$^{1,2}$, Minfeng Gu$^{1}$\\
$^{1}$Key Laboratory for Research in Galaxies and Cosmology, Shanghai Astronomical\\
         Observatory, Chinese Academy of Sciences, 80 Nandan Road Shanghai 200030, China \\
$^{2}$ University of Chinese Academy of Sciences, 19A Yuquanlu, Beijing 100049, China;
hxguo@shao.ac.cn, gumf@shao.ac.cn\\
} \shorttitle{The variability of SDSS quasars with multi-epoch spectroscopy} \shortauthors{Guo, et al.}

\begin{abstract}
In a sample of 60 quasars selected from Sloan Digital Sky Survey (SDSS) with at least six-epoch spectroscopy, we investigate the variability of emission lines and continuum luminosity at various aspects. A strong anti-correlation between the variability and continuum luminosity at 2500 {\AA} is found for the sample, which is consistent with previous works. In individual sources, we find that half of the sample objects
have a trend of being bluer-when-brighter trend (BWB), while the remaining half exhibit redder-when-brighter trend (RWB). Although the mechanism for RWB is unclear, the effects of host galaxy contribution due to seeing variations can not be completely ruled out. As expected from photoionization model, the positive correlations between the broad emission line and continuum luminosity are found in most individual sources, as well as for the whole sample. We confirm the Baldwin effect in most individual objects and the whole sample, while a negative Baldwin effect is also found in several quasars, which can be at least partly (if not all) due to the host galaxy contamination. We find positive correlations between the broad emission line luminosity and line width in most individual quasars, as well as the whole sample, implying a more variable line base than the line core.

\end{abstract}

\section{Introduction}
Active Galactic Nuclei (AGNs) are characterized by variability at almost all wavelengths \citep{ulr97}. Investigating the variability is a very important approach to probe the physical properties of AGNs.
There have been many systematic studies on the variability of AGNs by using photometric data, although such measurements are subject to the line contributions in the photometric bands (e.g. Vanden Berk et al. 2004). Several interesting results have been obtained from various AGN samples. The well-known anti-correlation between the variability and continuum luminosity was firstly discovered by \cite{Angione(1972)}, and later, it was confirmed by many other works \citep{Hook et al.(1994),Wilhite et al.(2008),zuo12}. Although there are other possibilities for explaining this result \citep{Cid Fernandes et al.(2000)}, \cite{Li(2008)} explained that the anti-correlation could be qualitatively explained by the standard accretion disc model assuming the variability was caused by the change of accretion rate.

A positive correlation  was found between the variability and redshift \citep*[e.g.,][]{Trevese et al.(1994),Hook et al.(1994),Cid Fernandes et al.(1996)}, which was later confirmed in a sample of over 25,000 SDSS quasars \citep{Vanden Berk et al.(2004),zuo12}. The correlation can be caused by the variability - wavelength relation \citep{Cristiani et al.(1996)}, which is likely related with the variability mechanism \citep{Vanden Berk et al.(2004)}. The positive relation between the optical-UV variability and black hole mass was first reported by \cite{Wold et al.(2007)}, which can be explained by the fact that the more massive black holes were gasless, and produce larger flux variations because they do not have a steady inflow of gaseous fuel. Alternatively, \cite{Li(2008)} argued that this relation could be triggered by the change of the accretion rate in their accretion disc model.

Mostly because the spectroscopic observations are time-consuming, there are only a few investigations on the variability of AGNs from multi-epoch spectroscopy, which either focused on reverberation mapping analysis for extensively monitored sources \cite[e.g.,][]{kaspi2000}, or were based on spectroscopic observations at a few epochs \cite[e.g.,][]{Wilhite et al. (2005)}. However, multi-epoch spectroscopy has advantages compared to photometric data. The continuum can be well constrained by excluding the emission line contamination, and the spectral shape can be measured by carefully fitting the continuum on a wide wavelength coverage. Moreover, the line measurements can be obtained, which enable us to study the variability of emission lines.

There have been extensive investigations on the quasar spectral shape, Baldwin effect and line width \citep*[e.g.,][]{Fan et al.(1998),Gu(2011)a,Gu(2011)b,Gu(2011)b,Wu et al.(2005),zuo12,Baldwin(1977),Wills et al.(1993),Wilhite et al. (2005)}. The bluer-when-brighter trend (BWB) is very common in AGNs, however the redder-when-brighter trend (RWB) has also been found, for example, in a sample of 544 quasars with two-epoch spectroscopy \citep{Bian et al.(2012)}. The anti-correlation between the emission line equivalent width (EW) and the continuum luminosity, so-called the Baldwin Effect \cite[][]{Baldwin(1977)} was originally found in the broad emission lines in UV/optical band \cite[see][for a review]{Shields(2007)}, and was also  detected in narrow lines \citep*[e.g.,][]{Zhangk et al.(2013)}. Recently, a strong Baldwin effect for \civ and \mgii and a week negative Baldwin effect for \hb were presented by \cite{Shen et al.(2011)} for DR7 quasar catalog. There is still no definitive conclusion on the correlation between the emission line luminosity and the line width. A negative correlation was found by \cite{Wills et al.(1993)} in a sample of 123 quasars with single-epoch spectroscopy, while a positive correlation was discovered by \cite{Wilhite et al. (2005)} based on a sample of 315 quasars with two-epoch spectroscopy. We note that all these works focused on studies of entire samples; not much work has been done for individual quasars. Moreover, usually only two-epoch spectroscopy was used in individual sources. To further study the Baldwin effect, the variability of the spectral shape and line width, especially in individual objects, the quasar samples with multi-epoch spectroscopy data are needed.

In this paper, we investigate the quasar variabilities by constructing a sample of quasars with multi-epoch spectroscopy from Sloan Digital Sky Survey (SDSS) \footnote{\tt http://dr9.sdss3.org/bulkSpectra}  \citep{abazajian2009,ahn12}. The multi-epoch spectroscopic data enables us not only to study the continuum variability, but also the effects that involve line variations relative to the continuum and relative to the line width, for both individual quasars and whole sample. In order to increase the probability of detecting variabilities, and to improve correlation analysis for individual QSOs, the quasars with at least six-epoch spectroscopy were selected. In Section 2, we describe the quasar sample, and the spectroscopic data analysis is given in Section 3. We show the results, and discussions in Sections 4, and 5, respectively. Finally, our conclusions are drawn in Section 6. Throughout the paper, a cosmology with $H_{0}=70\rm ~km~ s^{-1}~Mpc^{-1}$, $\Omega_{\rm m}=0.3$ and $\Omega_{\Lambda}=0.7$ is adopted, and the spectral index $\alpha_{\lambda}$ is defined as $f_{\lambda}$ $\propto$ $\lambda^{{\alpha}_{\lambda}}$ with $f_{\lambda}$ being the flux density at wavelength $\lambda$.

\section{Sample}

The SDSS data release seven (DR7) quasar catalog consists of 105,783 quasars selected as brighter than ${M_{\rm i}}$ = -22.0 and to have at least one broad emission line with FWHM larger than 1000 km $\rm s^{-1}$ \citep{sch10}. These quasars were selected for spectroscopic observation according to the quasar target selection algorithm \citep*[][]{Richards2002,sch10}, which selects objects with $i<19.1$ (for $i$-band apparent magnitude $i$) and their nonstellar colors similar to redshift $\le 3$ quasars, and unresolved objects with $i<20.2$ and colors similar to higher-redshift quasars. Moreover, all $15<i<19.1$ unresolved sources within 2" of a FIRST radio detection were also chosen.

The spectral wavelength coverage is 3,800{\AA} - 9,200{\AA} with the spectral resolution R $\sim$ 1,850 - 2,200, and the five-band $\emph {u,g,r,i,z}$ magnitudes have typical errors of about 0.03 mag. A comprehensive compilation of quasar properties is presented for DR7 quasars in \cite{Shen et al.(2011)}, including the continuum and emission line measurements, black hole masses, and radio properties etc.. To study the variabilities, the quasars with at least two-epoch spectroscopic observations were selected by searching the number of spectroscopic observations given in \cite{Shen et al.(2011)}, which results in a sample of 7,063 quasars. The multi-epoch spectroscopic observations are mainly from the overlap survey areas between adjacent plates, and sometimes were used to monitor the system \citep*[][]{dawson2013}. As the first of a series papers, we present in this work the results of 60 quasars with at least six-epoch spectroscopic observations. The redshift covers from 0.08 to 3.78 for these 60 objects. Eight quasars have been detected in the Faint Images of the Radio Sky at Twenty Centimeters (FIRST) 1.4-GHz radio survey \citep{bec95}. The radio loudness is available for all eight sources in \cite{Shen et al.(2011)}, of which seven objects are radio-loud according to the definition of radio loudness $R=f_{\rm 6 cm}/f_{\rm 2500 \AA} \ge 10$ ($f_{\rm 6 cm}$ and $f_{\rm 2500 \AA}$ are the flux density at rest-frame 6 cm and 2500 \AA, respectively). When available, we include in our analysis the spectra from SDSS data release nine (DR9), which is the first spectroscopic data from the SDSS-III Baryon Oscillation Spectroscopic Survey (BOSS) \citep{ahn12}. The spectra of DR9 cover a wider wavelength range 3,600 - 10,500 {\AA} than that of DR7. The biggest advantage of our sample is that each of our sample sources have at least six-epoch spectroscopy, therefore, we will mainly focus on the results for individual objects, instead of global sample as our sample size is relatively small compared to previous works.

Since different emission lines are covered in SDSS spectra for sources at different redshift, our sample sources are separately listed in Tables \ref{table:1} - \ref{table:3}. Table \ref{table:1} lists the quasars at $z\le0.4$, while the objects at $0.4<z\le0.8$, and $z>0.8$ are presented in Tables \ref{table:2} and \ref{table:3}, respectively.

\section{Spectroscopic analysis}

\begin{figure*}
\centering
\includegraphics[]{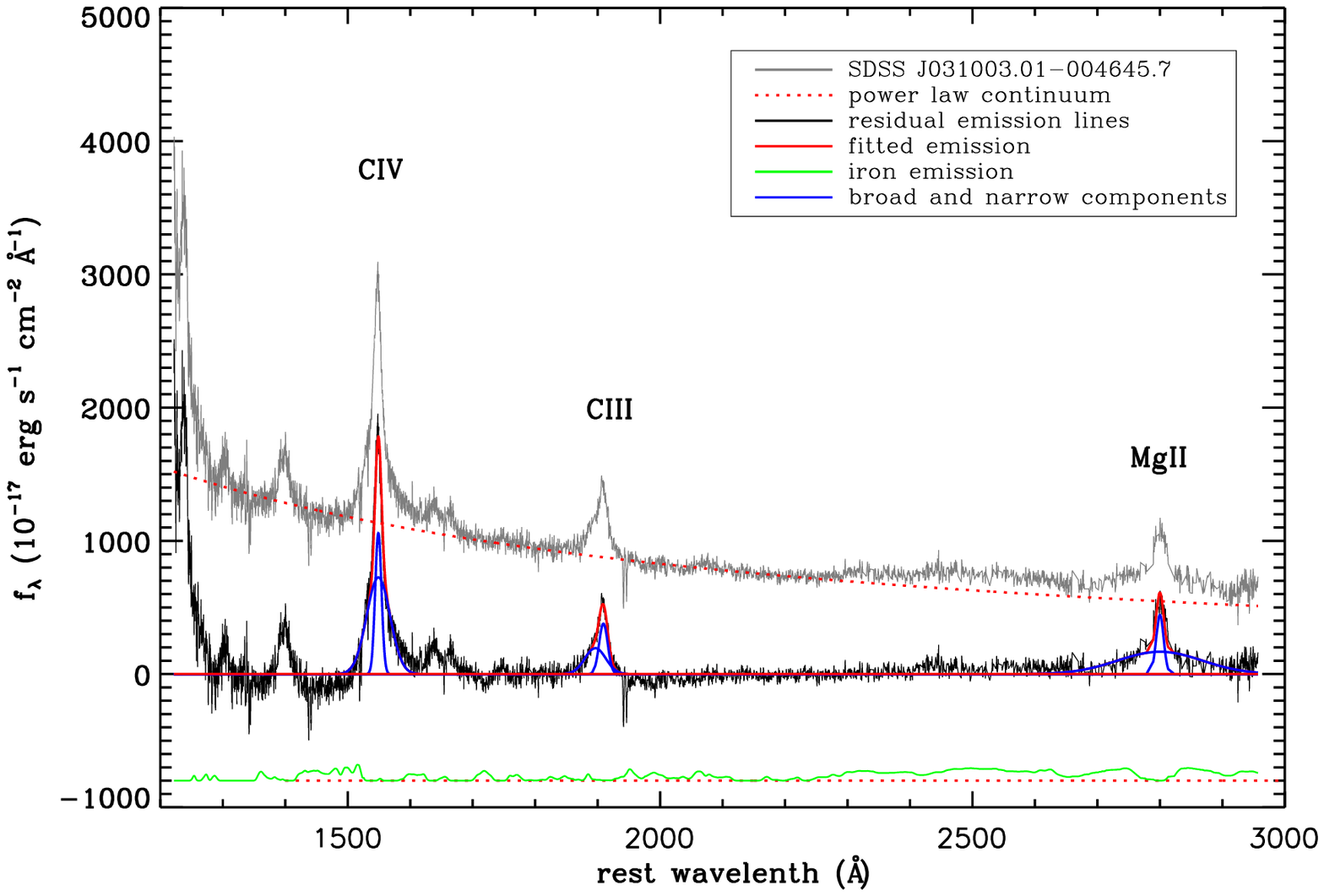}
\caption{The spectral analysis of SDSS J031003.01-004645.7 as an example. The grey and black lines are the original and continuum-subtracted spectrum, respectively. The red dotted line represents the power-law continuum. The blue lines are each Gaussian component of various emission lines, and the red solid line is the integrated profile. The green line is Fe II emission, which is shifted downwards with arbitrary unit for the sake of presentation. The prominent emission lines, \civ, \ciii, and \mgii are marked.}

\label{fig:belowzero}
\end{figure*}

The data reduction on the SDSS spectra follows the procedure in Chen et al. (2009), which is illustrated in Fig. \ref{fig:belowzero}.
After correcting the Galactic extinction
with the reddening map of \cite{Schlegel et al.(1998)}, we
shifted the spectra to the rest-frame wavelength. In order to obtain reliable line parameters, line-free wavelength
ranges were firstly selected as pseudo-continua. In addition to the emission lines, three components are considered: (1) A power-law continuum derived from the emission line-free windows;
 (2) UV and optical Fe II emission fitted using the templates of \cite{Vestergaard(2001)} and \cite{veron-cetty2004}, respectively; (3) A Balmer continuum generated in
the same way as \cite{Dietrich et al.(2002)}. The modeling of these three components was performed by minimizing
the $\chi^2$ in the fitting process. The final multicomponent fit was
then subtracted from the observed spectrum.

The broad emission lines were measured from the continuum subtracted
spectra. We mainly focus on several prominent emission lines, i.e.,
\ha, \hb, \mgii, and \civ. The \mgii, \hb and \ha lines were fitted with two Gaussian components, with one for narrow component with an upper limit of FWHM $\leq$ 1200 km $\rm s^{-1}$, and the other for broad profile with a lower limit of FWHM $\geq$1200 km $\rm s^{-1}$ \citep*[see e.g.,][]{Shen et al.(2011)}. Although it is still unclear whether there is a narrow component, we fitted \civ with two Gaussians.

The blended narrow lines, e.g., [O III] $\lambda\lambda4959,5007\rm \AA$
and [He II] $\rm \lambda 4686 \AA$ blending with \hb, and
[S II] $\lambda\lambda6716,6730\rm \AA$, [N II]
$\lambda\lambda6548,6583\rm \AA$ and [O I]$\rm \lambda 6300 \AA$
blending with \ha, are included as one Gaussian component
for each line at the fixed line wavelength. The details of spectral analysis were given in \cite{chen 2009}. The spectral index of the continuum, the continuum flux, the line width and flux
of broad \ha, \hb, \mgii and \civ lines were obtained from the final fits for our sample, from which the variations of both the continuum and line emission can be investigated for individual sources and whole sample.

\section{Results}

\subsection{The variability in continuum luminosity}

\begin{figure*}
\centering
\includegraphics[]{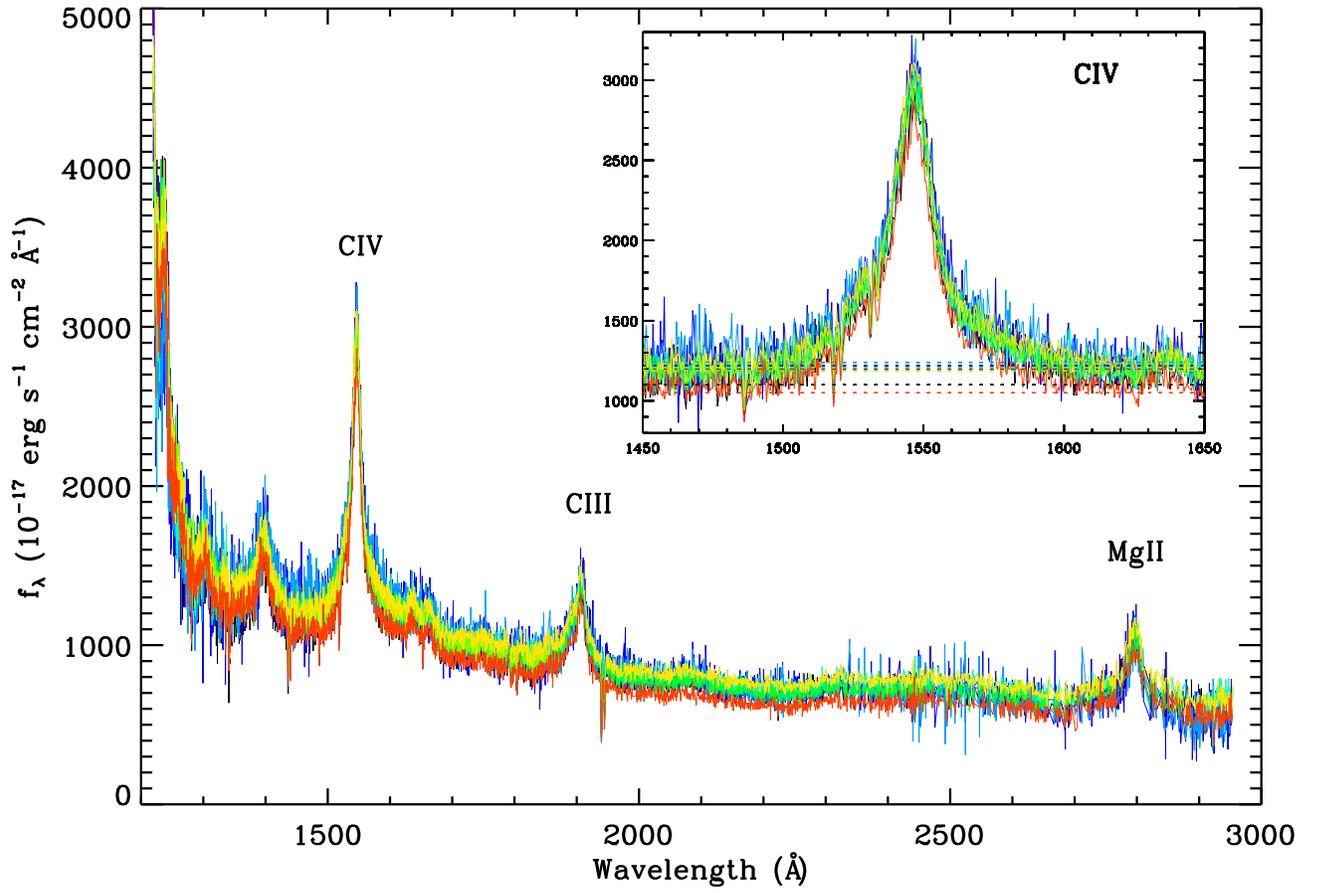}
\caption{The nine-epoch SDSS spectra of SDSS J031003.01-004645.7 ($z=2.115$) in the rest frame. Each epoch is indicated with different color. The dashed lines in the inset are the local continuum for \civ. }
\label{fig:zero}
\end{figure*}

The variability of continuum emission is investigated with the rest luminosity at 2500 {\AA}. This wavelength is selected because it is covered in the SDSS spectra for most sources, and the variations at shorter wavelength are known to be larger than longer wavelength \citep*[e.g.,][]{Cristiani et al.(1997)}. The luminosity at 2500 {\AA} is directly calculated from the fitted power-law continuum if 2500 {\AA} is covered in the spectrum, otherwise is extrapolated from the power-law continuum, which is the case in seven quasars. We define the continuum variability amplitude for each source as $\Delta \rm log~\it \lambda L_{\lambda} = \rm log ~\lambda \it L_{\rm \lambda,max} - \rm log~\lambda L_{\rm \lambda,min}$, in which $\lambda L_{\rm \lambda,max}$, and $\lambda L_{\rm \lambda,min}$ are the highest and lowest luminosity at 2500 {\AA} measured from multi-epoch spectra, respectively. To evaluate the significance of variability, two uncertainties are taken into account. The first one is the measurement uncertainties in two involved spectra $\sigma_{\rm s} = \sqrt[]{ \sigma^2_{\rm s,1} + \sigma^2_{\rm s,2}}$, in which $\sigma_{\rm s,1}$, and $\sigma_{\rm s,2}$ are uncertainties in two spectra. The other is the uncertainties from the power-law fitting $\sigma_{\rm f} = \sqrt[]{ \sigma^2_{\rm f,1} + \sigma^2_{\rm f,2}}$. The total uncertainty in $\Delta \rm log~\it \lambda L_{\lambda}$ is $\sigma = \sqrt[]{\sigma^2_{\rm s} + \sigma^2_{\rm f} }$. We found that  $\Delta \rm log~\it \lambda L_{\lambda}$ is larger than $3\sigma$ for all our sources, implying significant continuum variations in our sample. As an example, the continuum variation is shown in Fig. \ref{fig:zero} for nine-epoch spectra of SDSS J031003.01-004645.7 ($z=2.115$), in which the \civ line variation is also presented.

\begin{figure*}
\centering
\includegraphics[]{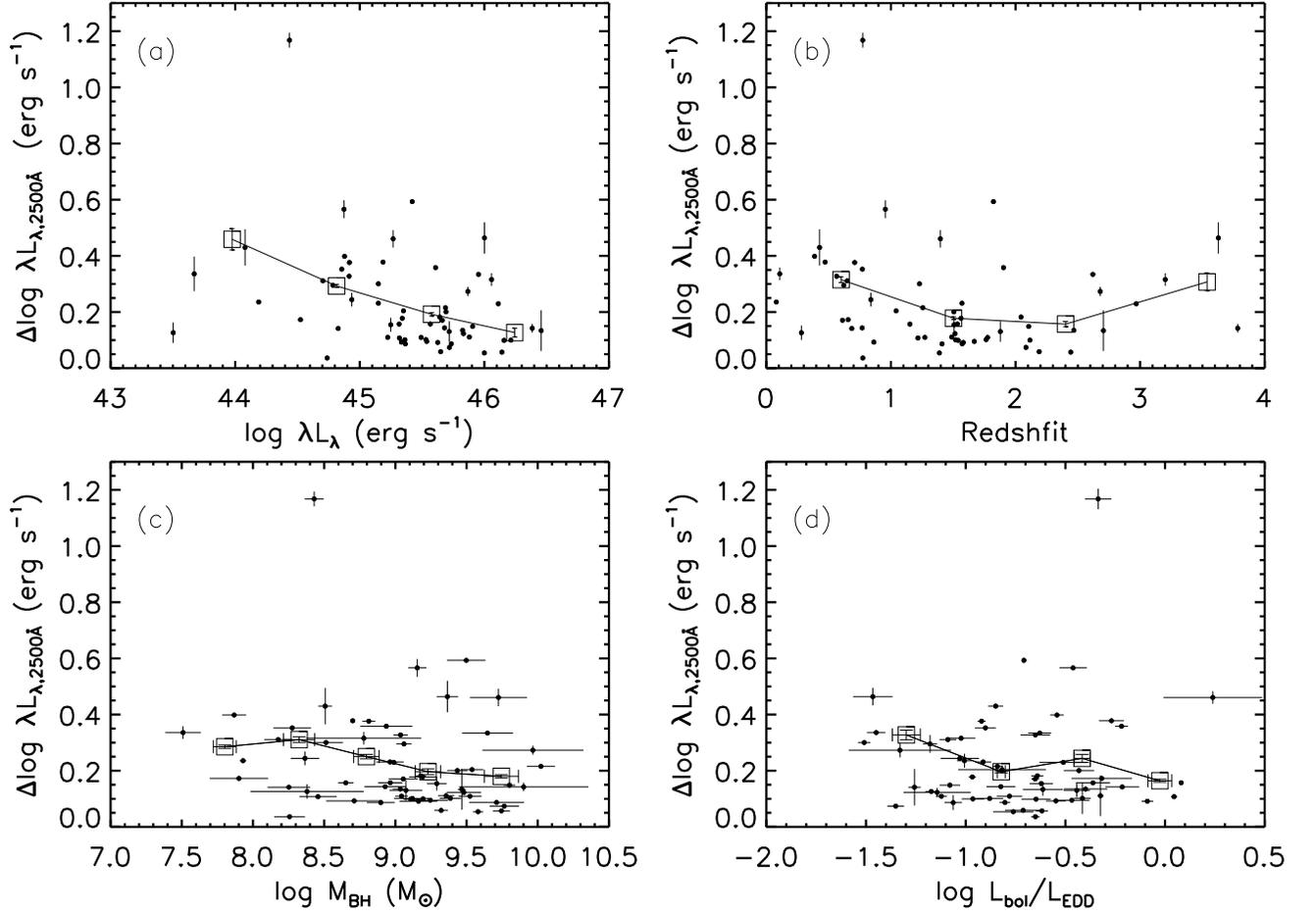}
\caption{(a): The variability amplitude versus continuum luminosity at 2500\AA. The continuum luminosity is the multi-epoch averaged value; (b): The variability amplitude versus redshift; (c): The variability amplitude versus black hole masses; (d): The variability amplitude versus Eddington ratio. In each panel, the squares represent the mean variability amplitude in bins of $x$-axis parameters. The different bins are divided by (a) 44.5, 45, 46; (b) 1, 2, 3; (c) 8.0, 8.5, 9.0, 9.5; (d) -1.1, -0.6, -0.2.
}
\label{fig:four}
\end{figure*}

The relationship between the continuum variability amplitude and various parameters are presented in Fig. \ref{fig:four} for our sample. We find a significant anti-correlation between the continuum variability amplitude and the multi-epoch averaged 2500 {\AA} luminosity with a Spearman rank correlation coefficient $r_{\rm s} = -0.38$ at $\sim99.7\%$ confidence level (see Fig. \ref{fig:four}a). This anti-correlation is more apparent in binned 2500 {\AA} luminosity.
In  Fig. \ref{fig:four}b, we find a mild anti-correlation between the redshift and the continuum variability amplitude with a Spearman correlation coefficient $r_{\rm s} = -0.27$ at $\sim96.6\%$ confidence level. In redshift bins, the continuum variability amplitude decreases with the redshift at $z<2.5$, while it increases at higher redshift, where however there are only a few objects. For our objects, the variability was obtained from the luminosity at the same rest frame wavelength 2500 {\AA}, which naturally eliminates the selection effect of rest wavelength. However, our result is in contrast to the positive correlations reported in previous works \citep*[e.g.,][]{Cristiani et al.(1996),Vanden Berk et al.(2004)}.

By collecting the black hole mass $M_{\rm bh}$ and the Eddington ratio $L_{\rm bol}/L_{\rm Edd}$ from \cite{Shen et al.(2011)} (see Tables \ref{table:1} - \ref{table:3}), their relations with the continuum variability amplitude were studied for our sample (see Figs. \ref{fig:four}c and \ref{fig:four}d). From the Spearman correlation analysis, we failed to find the significant correlation between the continuum variability amplitude and the black hole mass. Similarly, there is no strong correlation between the continuum variability amplitude and the Eddington ratio, although the trend of decreasing variability amplitude with increasing $M_{\rm bh}$ and $L_{\rm bol}/L_{\rm Edd}$ can be seen from the binned values.

In Fig. \ref{fig:four}, it can be clearly seen that one quasar (SDSS J022214.38-001745.3, $z=0.773$) has much larger variability amplitude  than all other objects. In order to evaluate its influence on the correlations, we performed the correlation analysis after excluding the object, and found similar correlation results.

\subsection{The spectral variation}

\begin{figure*}
\centering
\includegraphics[]{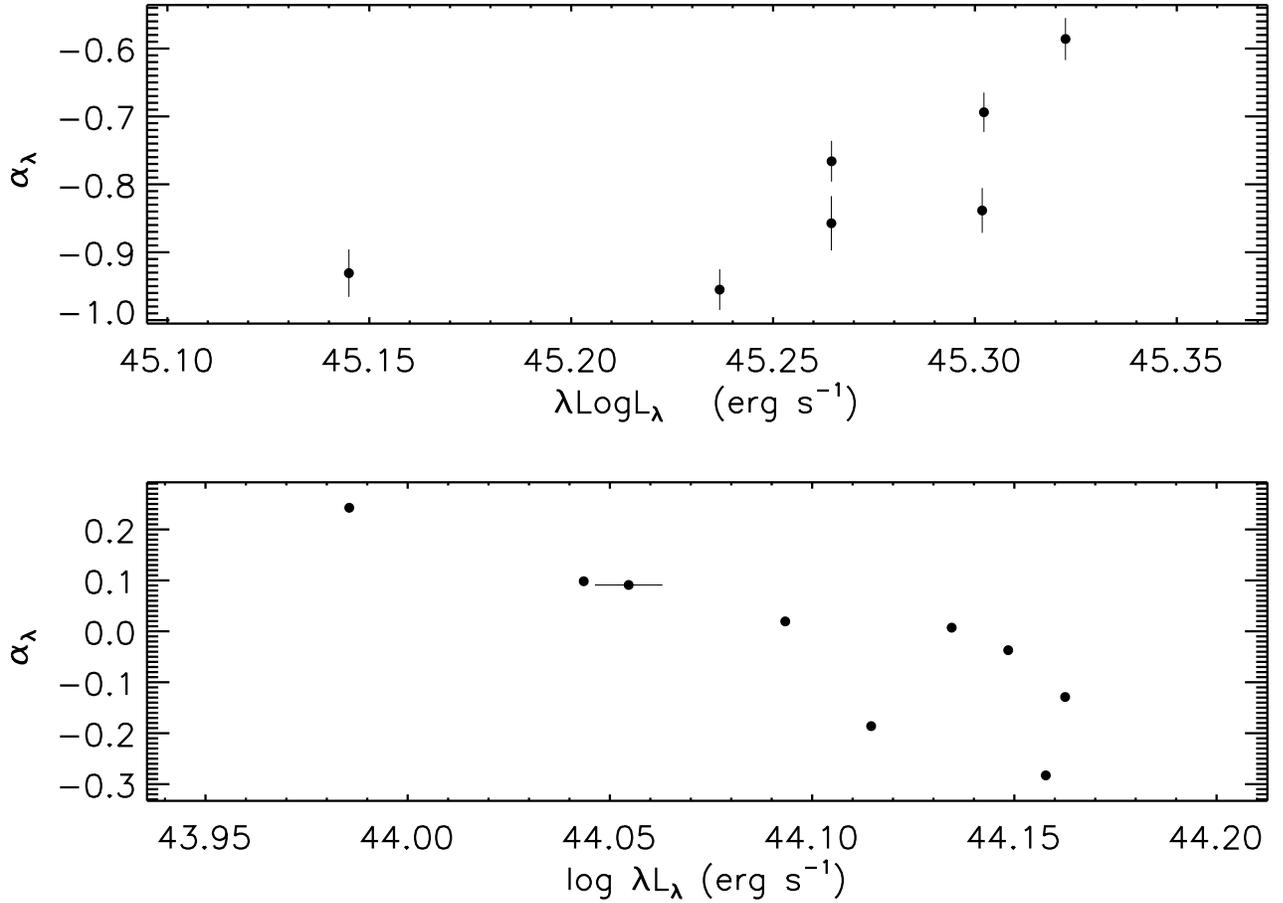}
\caption{The examples of correlations between the spectral index and the continuum luminosity at 2500 \AA: RWB (SDSS J031022.10+004130.0, top), and BWB (SDSS J30639.57+000343.1, bottom). The errors of the spectral index and continuum luminosity are indicated by the vertical and horizontal lines, respectively, which are not evident when less than the symbol size.}
\label{fig:indexgood}
\end{figure*}

The multi-epoch spectroscopic observations enable us to investigate the relationship between the 2500 {\AA} continuum luminosity and the spectral shape ($\alpha_{\lambda}$) for individual sources. The Spearman correlation coefficient and the confidence level are shown for each source in Tables \ref{table:1} - \ref{table:3}.
We find that 30 sources have RWB trend with positive correlations, while the rest of the 30 objects show BWB trend with negative correlation coefficient. While the correlation confidence level is quite low for most objects, we find significant anti-correlations in three quasars at confidence level of $\geq$ 99$\%$, and mild anti-correlations in six objects at $90\%\leq$ p $\leq 99\%$ confidence level. In contrast, none of sources have significant positive correlations, and only three quasars have mild positive correlations (SDSS J022518.36-001332.3 at $z = 3.63$, SDSS J030907.49+002419.0 at $z = 2.08$, and SDSS J031022.10+004130.0 at $z = 0.65$, see Tables \ref{table:2} - \ref{table:3}). The correlations are illustrated in Fig. \ref{fig:indexgood} with two examples, SDSS J030639.57+000343.1 (BWB, $r_{\rm s}=-0.89$ at $>99.5\%$), and SDSS J031022.10+004130.0 (RWB, $r_{\rm s}=0.67$ at $\sim91\%$).

\subsection{Line emission}

The correlation between the variations in the broad emission line luminosity with the continuum luminosity is explored for each object, and the results from Spearman correlation analysis are shown in Tables \ref{table:1} - \ref{table:3}, in which the continuum luminosity at 1350 {\AA}, 3000 {\AA}, 5100 {\AA}, and 5100 {\AA}, correspond to broad \civ, \mgii, \hb, and \ha lines, respectively. 

\begin{figure*}
\centering
\includegraphics[]{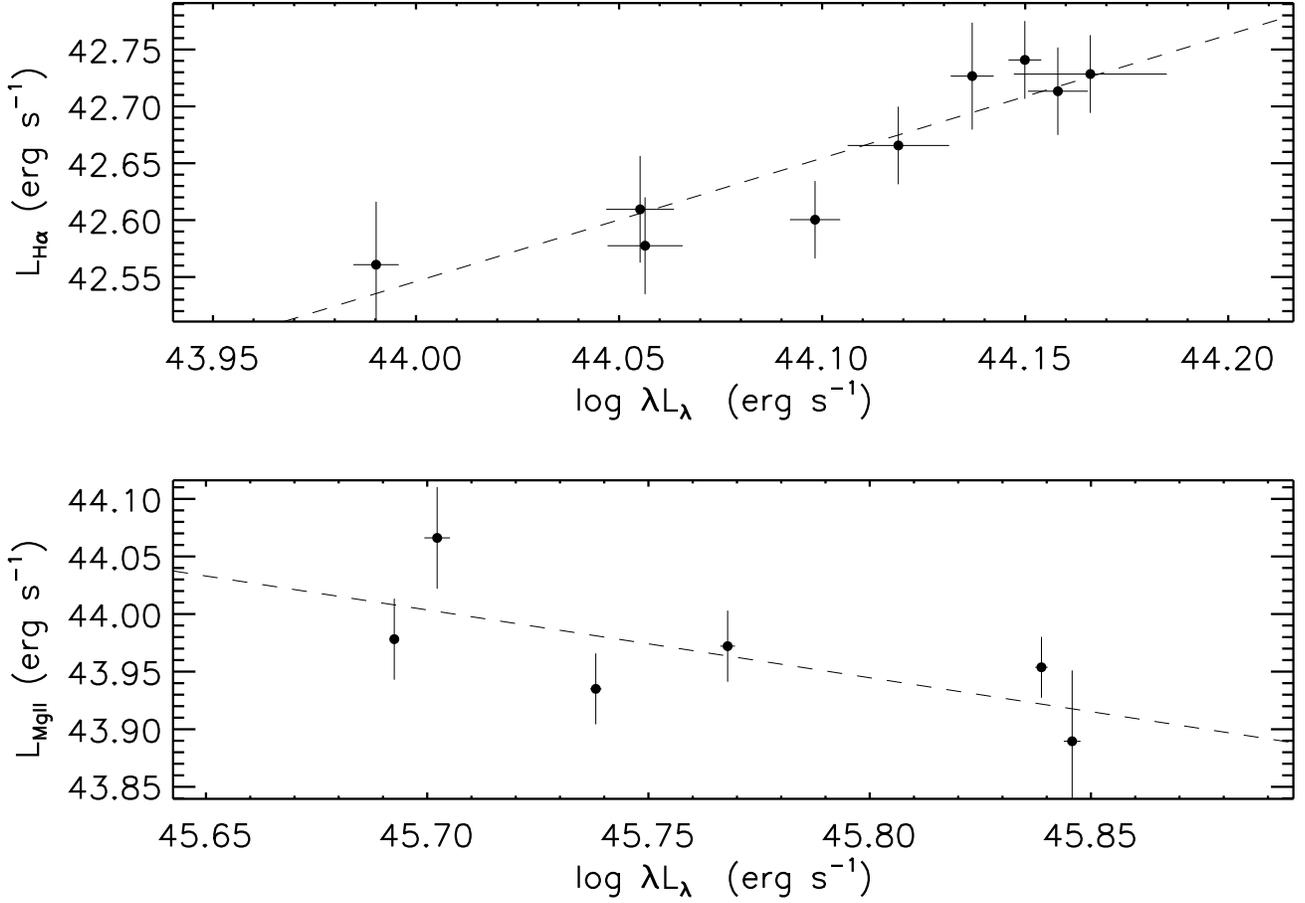}
\caption{The examples of the positive correlation (SDSS J030639.57-000343.1, for \ha, top) and anti-correlation (SDSS J021754.80+000234.0, for \mgii, bottom) between the broad emission line and continuum luminosity. The dashed lines are linear fits (see text for details). The errors of the broad emission line and continuum luminosity are indicated by the vertical and horizontal lines, respectively, which are not evident when less than the symbol size.}
\label{fig:LLgood}
\end{figure*}

\begin{figure*}
\centering
\includegraphics[]{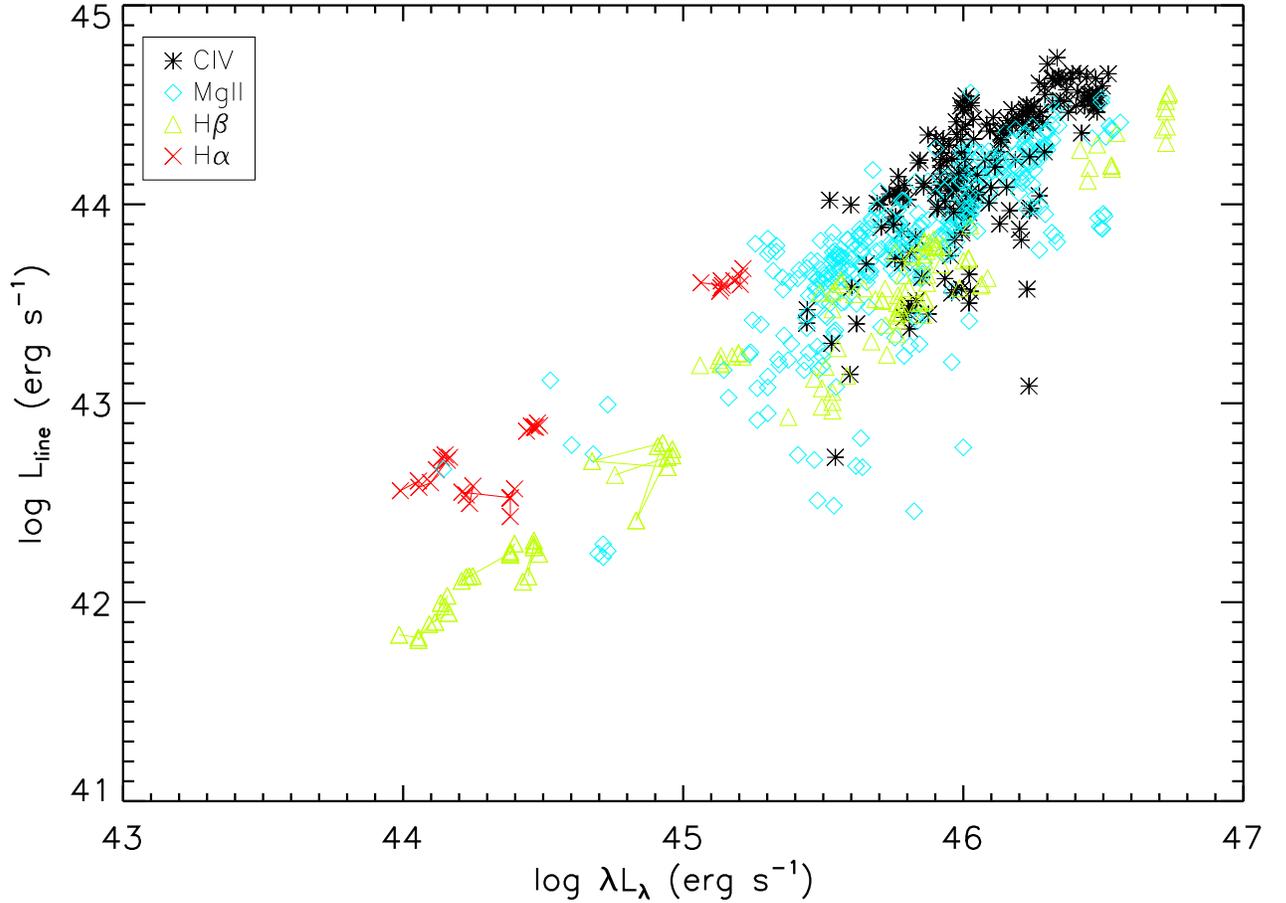}
\caption{The broad emission line and continuum luminosity from all multi-epoch spectra of our sample. The continuum luminosity for broad \civ, \mgii, \hb, and \ha are at 1350{\AA}, 3000{\AA}, 5100{\AA}, and 5100{\AA}, respectively. The measurements from multi-epoch spectra for the same object are connected with solid lines for \ha in four sources, and \hb in four sources.}
\label{fig:allLL}
\end{figure*}

We find positive correlations between the line and continuum luminosity in most cases (74 of 92 emission lines, $\sim80\%$), consistent with the photoionization model (Yee 1980), of which $\sim26\%$ (24 of 92) have $\geq 90\%$ correlation confidence level. As an example, the correlation between the broad \ha luminosity and the continuum luminosity at 5100 {\AA} is plotted in Fig. \ref{fig:LLgood} for SDSS J030639.57+000343.1. A significant correlation is found with $r_{\rm s} = 0.867$ at confidence level of 99.7$\%$ (see also Guo \& Gu 2014). The linear fit gives
\begin{equation}
\rm log~\it L_{\rm H\alpha} =\rm (1.03 \pm 0.13)~log~(\it \lambda L_{\rm \lambda,{5100\AA}}) \rm - (2.83 \pm 5.81).
\end{equation}
Interestingly, we find anti-correlations between the line and continuum luminosity in the rest of the cases. However, the anti-correlations are usually very weak (see Tables \ref{table:1} - \ref{table:3}). Only one object (SDSS J021754.80 +000234.0) exhibits a mild anti-correlation with $r_{\rm s} = -0.77$ at $\sim93.0\%$ confidence level (see Table \ref{table:3}), which is shown in Fig. \ref{fig:LLgood} for the broad \mgii luminosity and the continuum luminosity at 3000 {\AA}. The linear fit yields
\begin{equation}
\rm log~\it L_{\rm MgII} =\rm  (-0.50 \pm 0.32)~log~(\it \lambda L_{\rm \lambda,{3000\AA}}) \rm + (70.91 \pm 14.66).
\end{equation}
When putting all multi-epoch line measurements of all our sample sources together, we find significant positive correlations between each line emission and the corresponding continuum luminosity for whole sample (see Fig. \ref{fig:allLL}).

\subsection{Baldwin effect}
\begin{figure*}
\centering
\includegraphics[]{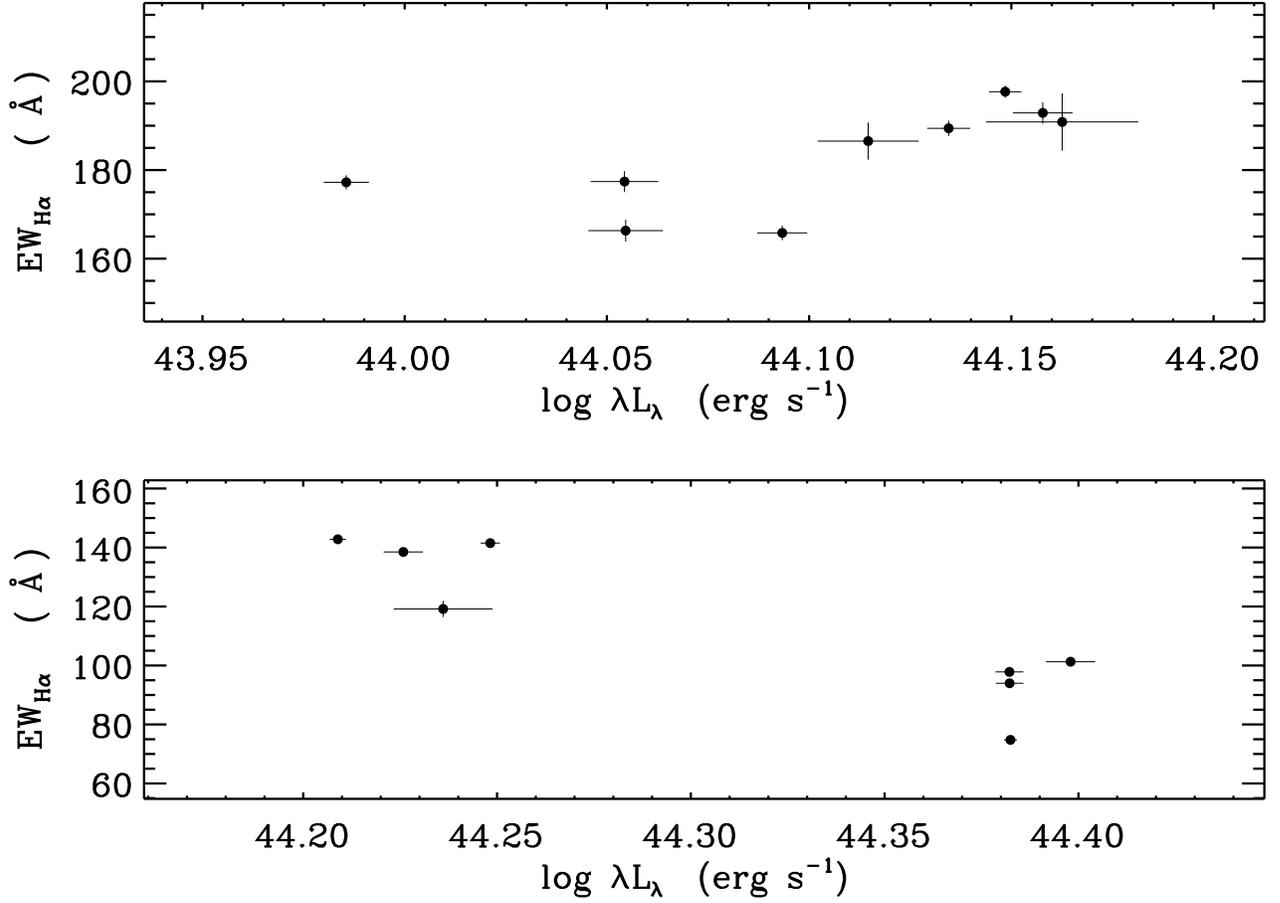}
\caption{The examples of the positive correlation (SDSS J030639.57+000343.1, for \ha, top) and negative correlation (SDSS J031027.82-004950.7, for \ha, bottom) between the broad line equivalent width and continuum luminosity. The errors of the EW and continuum luminosity are indicated by the vertical and horizontal lines, respectively, which are not evident when less than the symbol size.}
\label{fig:baldgood}
\end{figure*}

The measurements of emission lines from multi-epoch spectra also enable us to study the Baldwin effect for both individual sources and whole sample. In Tables \ref{table:1} - \ref{table:3}, we list the Spearman correlation coefficient and corresponding confidence level between the EW (broad \civ, \mgii, \hb, and \ha) and the continuum luminosity for individual sources. We find that 61 of 92 ($\sim66\%$) emission lines show anti-correlations, with 13 cases ($\sim14\%$) at confidence level of $\geq 90\%$. Two sources are found to have significant anti-correlations at confidence level $>99\%$, i.e., SDSS J031226.12-003708.9 and SDSS J031156.45-004157.0 (see Tables \ref{table:2} and \ref{table:3}). Strikingly, the positive correlations, namely the negative Baldwin effect, are also found in 31 cases, however, the correlations are usually weak (confidence level $<90\%$). The examples of anti-correlation and positive correlation are shown in Fig. \ref{fig:baldgood} for SDSS J031027.82-004950.7 and SDSS J030639.57+000343.1, respectively.

\begin{figure*}
\centering
\includegraphics[]{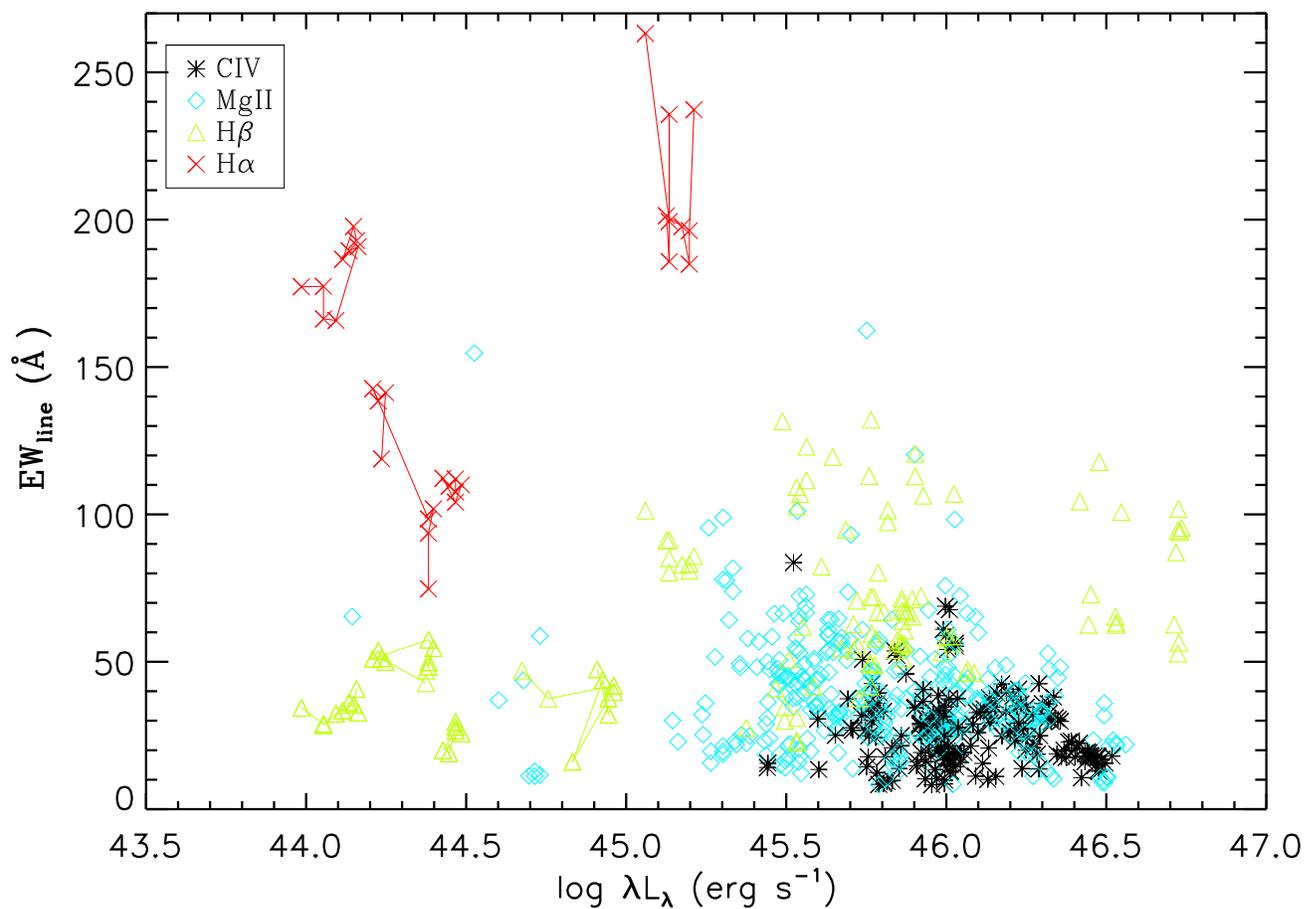}
\caption{The broad emission line equivalent width and continuum luminosity from all multi-epoch spectra of our sample. The measurements from multi-epoch spectra for the same object are connected with solid lines for \ha in four sources, and \hb in four sources.}
\label{fig:allbald}
\end{figure*}

We further study the Baldwin effect for whole sample by plotting all line EW measurements and the corresponding continuum luminosity in Fig. \ref{fig:allbald}. Although the scatter is large, a significant Baldwin effect, i.e., the strong anti-correlation between the line EW and continuum luminosity is found with a Spearman correlation coefficient of $r_{\rm s}=-0.23$ at $\gg 99.9\%$ confidence level for the broad \mgii line, which is covered in SDSS spectra for most sources.

\subsection{Line emission and width}

\begin{figure*}
\centering
\includegraphics[]{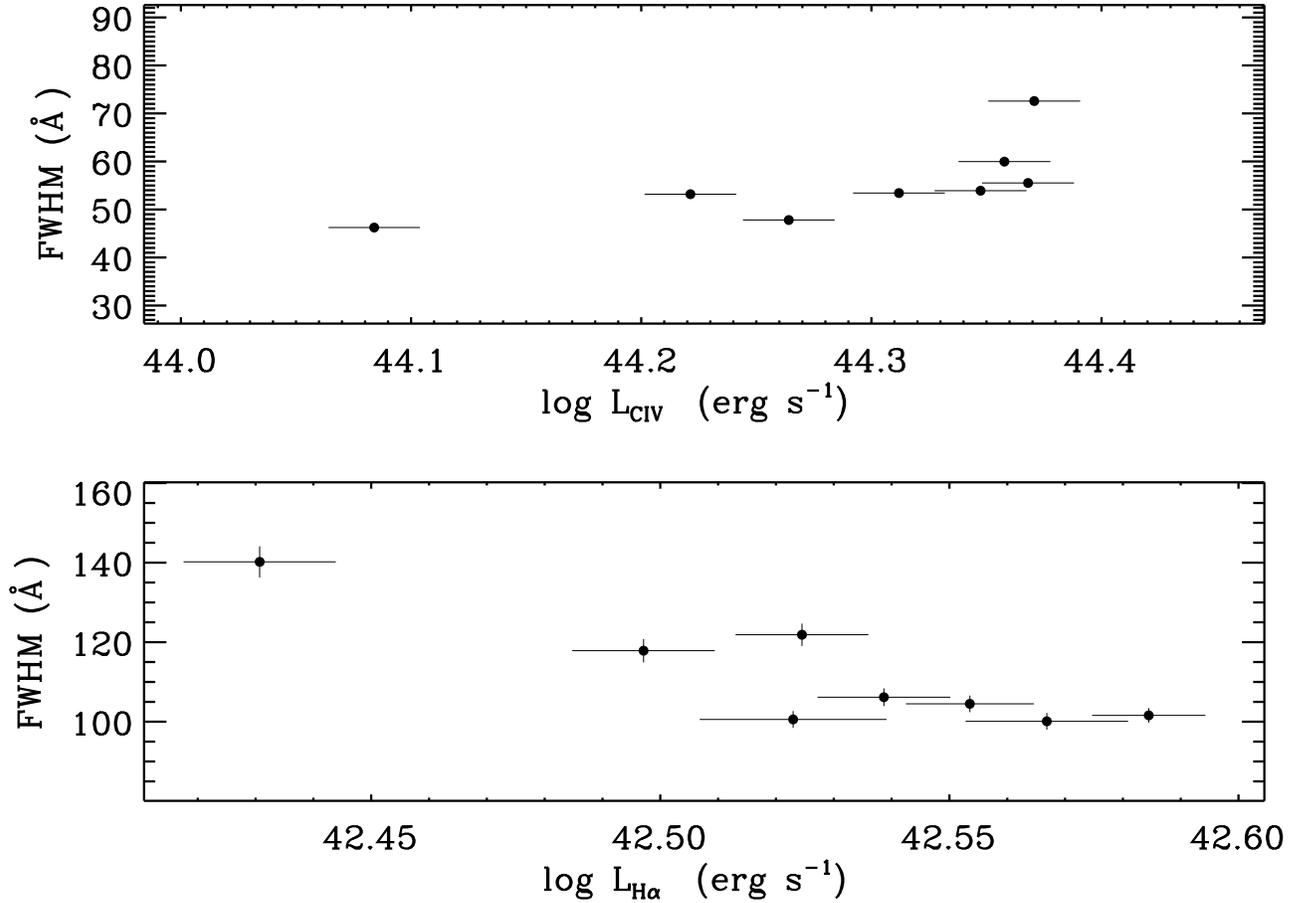}
\caption{The examples of the positive correlation (SDSS J031131.41-002127.4, for \civ, top) and negative correlation (SDSS J031027.82-004950.7, for \ha, bottom) between the FWHM and luminosity of broad lines. The errors of the FWHM and line luminosity are indicated by the vertical and horizontal lines, respectively, which are not evident when less than the symbol size.}
\label{fig:FWHMgood}
\end{figure*}

With multi-epoch spectroscopy, we investigate the relationship between the variability of the broad line width FWHM and the broad line luminosity (broad \civ, \mgii, \hb, and \ha) for individual sources. In Tables \ref{table:1} - \ref{table:3}, we show the Spearman correlation coefficient and confidence level for various emission lines in each source. While 69 of 92 ($\sim75\%$) emission lines show a positive correlation, with 24 cases ($\sim26\%$) at confidence level of $\geq 90\%$, the anti-correlations were also found. Whereas, anti-correlations are usually weak, with only three sources at $>90\%$ confidence level (\civ in SDSS J022230.28+001844.5, \mgii in SDSS J030911.64+002358.8, and \ha in SDSS J031027.82-004950.7, see Tables \ref{table:1} - \ref{table:3}). In Fig. \ref{fig:FWHMgood}, the positive and anti-correlation are illustrated with two examples, SDSS J031131.41-002127.4 and SDSS J031027.82-004950.7, for broad \civ, and \ha, respectively.

\begin{figure*}
\centering
\includegraphics[]{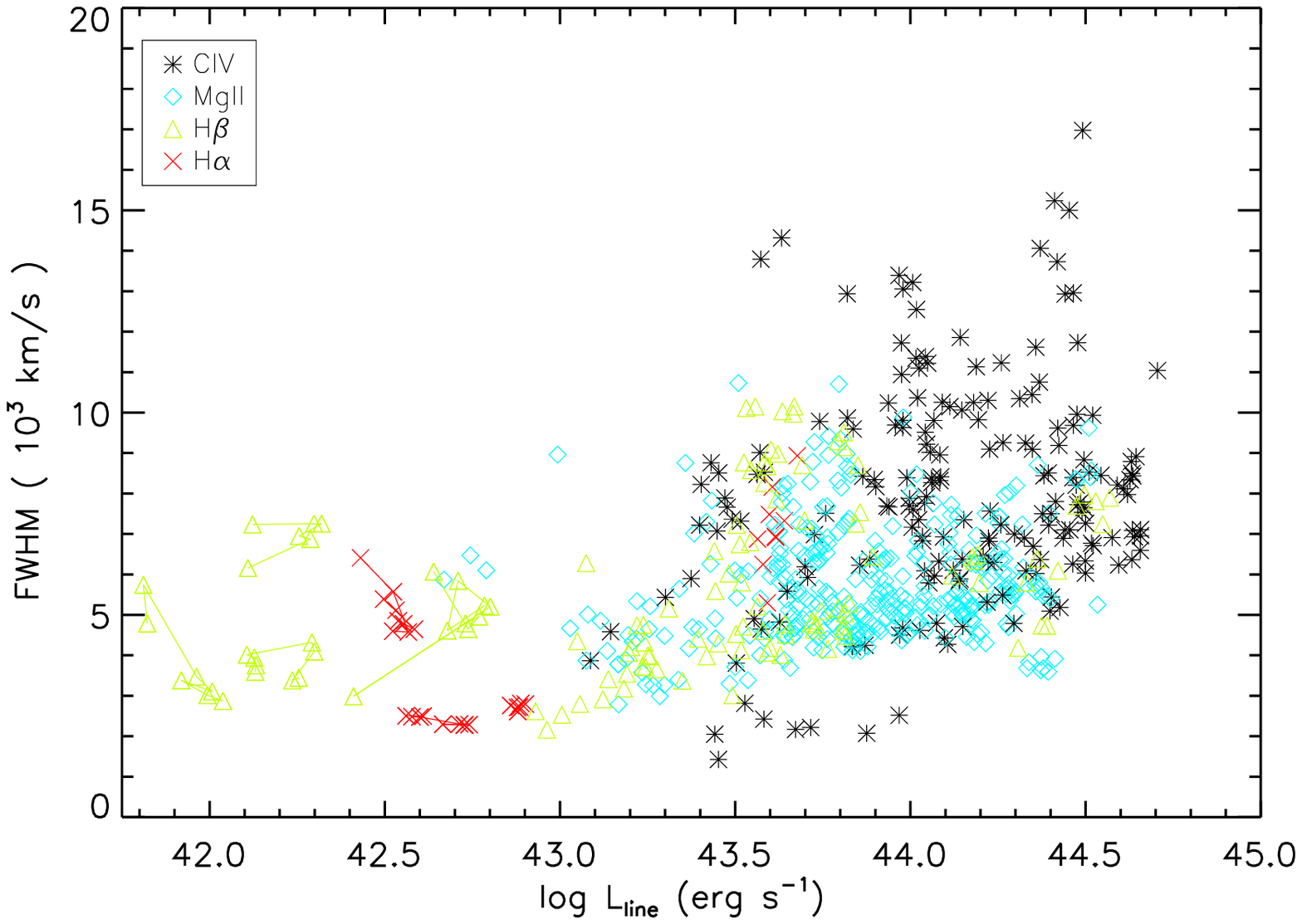}
\caption{The broad emission line FWHM and line luminosity from all multi-epoch spectra of our sample. The measurements from multi-epoch spectra for the same object are connected with solid lines for \ha in three sources, and \hb in four sources.}
\label{fig:allWHM}
\end{figure*}

The relationship between broad line width and luminosity is also explored for whole sample when putting all multi-epoch measurements together (see Fig. \ref{fig:allWHM}). The strong correlations are found with Spearman correlation coefficients of 0.14, 0.16, 0.42 and 0.41, all at $>99\%$ confidence level for \civ, \mgii, \hb and \ha, respectively. The correlation analysis is not committed for \ha, since only four sources have \ha measurements. In Fig. \ref{fig:allWHM}, irrespective of large scatters, the mean values of line FWHM imply that systematically \hb and \mgii are at similar region, while \civ are relatively closer to central nuclei, with the mean FWHM of 41 {\AA} ($\sim 7900~ \rm km~s^{-1}$), 48 {\AA} ($\sim 5100 ~\rm km~ s^{-1}$), and 90 {\AA} ($\sim 5500~\rm  km~s^{-1}$) for broad \civ, \mgii, and \hb, respectively.

\section{Discussions}
\subsection{The color variation}

As mentioned in Section 2, the quasars are selected for candidates for spectroscopic observations according to various criterions. While most sources are chosen from their color, some are selected as the detections in FIRST survey. We find that the mean spectral index of all sample $<\alpha_{\lambda}>=-1.37\pm0.18$ is generally consistent with that of the composite spectrum \cite[$\alpha_{\nu}=-0.44$,][]{van01}, which indicates no severe reddening for whole sample. There are 15 quasars with $i>19.1$ (see Tables 1-3), and their mean spectral index is $-1.34$, similar to the value of the rest 45 bright objects $-1.38$. This implies that the reddening is not strong in faint sources, consistent with the color selection criterion.  In contrast, we find that the mean spectral index of the eight FIRST detected quasars $-1.07$, is evidently redder than that of non-FIRST detections $-1.42$. This can be likely due to the higher reddening, or the contamination of synchrotron emission in FIRST sources. However, none of these FIRST sources exhibits significant positive correlation between the spectral index and continuum luminosity, indicating that the reddening or synchrotron emission may not be necessarily related with RWB trend.

The anti-correlation of the variability amplitude with rest wavelength has been found in quasars \citep*[e.g.,][]{Cristiani et al.(1997),Vanden Berk et al.(2004),zuo12}. The spectrum shortward of 2500{\AA} shows a steeper slope and is more variable, while the spectrum longward of 2500 {\AA} is flat and less variable relatively for quasars dominated by Balmer emission and Fe II emission lines. This variance mainly comes from the changing of the temperature of the accretion disk \citep*[][]{Bian et al.(2012)}. It is reasonable to assume that most variability occurs in the inner part of the AGNs, so when the accretion disk becomes hotter, it will produce more high energy photons and the continuum emission peak will move to the short wavelength, which yields a bluer spectra when AGNs become brighter \citep*[see][]{Bian et al.(2012)}, i.e. the commonly observed BWB trend in AGNs \citep*[e.g.,][]{Gu(2011)a,Gu(2011)b,zuo12}. Indeed, we find a BWB trend in half of our sample sources; however, we also find a RWB trend in the remaining half of the sources. In flat-spectrum radio quasars (FSRQs), \cite{gu06} proposed that the varying contribution of the thermal emission from the disk relative to the synchrotron jet emission can qualitatively explain the RWB trend. However, this scenario can not be used to explain three quasars with mild RWB correlations (see Section 4.2), because all of them are not detected in FIRST.

It is still unclear what causes the RWB trend in radio-quiet quasars \cite[e.g.,][]{Bian et al.(2012)}. The line contribution in a broad photometric band could somehow explain the RWB color variability in photometric data \citep{sch12,Wilhite et al. (2005)}; nevertheless, it is hardly applicable to our sample, since the continuum flux in a spectrum that resolves the quasar emission lines and there is no contamination from line emission. Alternatively, the variations of the contribution of the host galaxy could qualitatively produce the RWB trend in multi-epoch spectroscopy with a fixed fiber size (3 arcsec in SDSS-I/II) when seeing varies. The influence of variable seeing conditions on the observed variations has already been noticed in aperture photometry \cite[e.g.,][]{cel00}. Since luminous quasars are usually hosted in the bright elliptical galaxies, the host galaxies will be more extended in poor seeing conditions, then their contribution will be relatively smaller within a fixed aperture compared to light from the quasar. This will result in a RWB trend when seeing varies because the host galaxies are usually redder than quasars. We checked the correlations between the spectral index and seeing for three RWB sources with mild correlations. With available seeing, we indeed found correlations at similar confidence level as RWB correlations in two sources, SDSS J030907.49+002419.0 ($z=2.083$), and SDSS J031022.10+004130.0 ($z=0.656$). Therefore, the effects of host galaxy contribution cannot be completely ruled out, although it is hard to quantitatively evaluate this possibility, and the contribution of host galaxy may not be significant at rest frame wavelength for these two objects.

\subsubsection{Comparison with previous works}
There have been extensive investigations on the relationship between the spectral shape and continuum variability, especially in radio-loud AGNs \citep*[e.g.,][]{Fan et al.(1998),Gu(2011)a,Gu(2011)b}. Two trends of color variations have been found. The BWB trend is commonly found in blazars, as well as in radio quiet AGNs \citep*[e.g.,][]{Wu et al.(2005),Gu(2011)a,Gu(2011)b,zuo12}. However, the RWB trend has also been found \citep*[e.g.,][]{gu06,Bian et al.(2012)}, which can be caused by the contribution of thermal emission in case of flat-spectrum radio quasars \citep{gu06}. Based on the two-epoch spectroscopy for a sample of quasars, which consists of 312 radio-loud and 232 radio-quiet sources, \cite{Bian et al.(2012)} found that half of the objects show the RWB trend, and no obvious difference can be found between sub-samples of radio-quiet and radio-loud quasars. Therefore, our results are consistent with \cite{Bian et al.(2012)} in the fact that half of our quasars exhibit a RWB trend, and this RWB trend seems to have no relation with the radio detection. The results will be further investigated in a sample of about 2000 SDSS quasars with pronounced variations from multi-epoch spectroscopy (Guo \& Gu 2014, in preparation).

\subsection{Line emission}
While most sources show positive correlations between the broad line and continuum luminosity, some quasars exhibit anti-correlations, an effect not expected in photoionization model. This anti-correlation could be qualitatively explained by the time delay between the line and continuum variability, because the continuum emission needs a finite amount of time to reach the broad line region clouds. This is actually why reverberation mapping works for long-term monitoring AGNs \citep*[e.g.,][]{kaspi2000}. When we are observing the increased continuum emission, the observed line emission actually happens to correspond to the past declined phase due to the time delay. Therefore, an anti-correlation will be obtained.

\subsection{Baldwin effect}

Despite the fact that the Baldwin effect has been extensively studied, its origin is still unclear \citep*[e.g.,][]{Netzer et al.(1992),Dietrich et al.(2002),Baskin2004),Wu et al.(2009)}, and several causes of possible correlations (hence, possible physical causes) have been proposed: correlations with the black hole mass, the Eddington ratio and the luminosity \citep*[e.g.,][]{Bian(2004),Vestergaard(2006)}. Most likely, it could be explained by the trend that more luminous sources have softer Spectral Energy Distribution \citep{Netzer et al.(1992),Dietrich et al.(2002)}, which produce less ionizing photons.

While the Baldwin effect has been studied extensively for AGN samples \cite[e.g.,][]{bia12b}, few works have been done on the Baldwin effect in individual sources. With multi-epoch spectroscopy, we investigate the Baldwin effect for our sample objects. As described in Section 4, most sources exhibit an anti-correlation between EW and continuum luminosity, as expected from Baldwin effect, although the correlations are usually weak. As claimed in \cite{Wilhite et al. (2005)} the broad lines are less variable than the underlying continuum, resulting in a relationship known as the intrinsic Baldwin effect, which is intrinsic to each object \citep{Kinney et al.(1990)}. However, the intrinsic Baldwin effect in each object will be altered by the light-travel time effects in the broad-line region. Indeed, it has been shown that the scatter in the continuum-emission-line correlations is greatly reduced by removing light-travel time effects so that the emission-line flux are referred to the continuum that is driving them \citep*[e.g.,][]{pogge92}. Unfortunately, we are unable to remove the light-travel time effects for our quasars due to the limited spectroscopic data. Interestingly, we also found a strong positive correlation between the EW and continuum luminosity (i.e. negative Baldwin effect) in SDSS J030639.57+000343.1 (broad $\rm H\alpha$, see Fig. 7) at $\sim99\%$ confidence level. This however is most likely caused by increasing host contamination toward fainter luminosities \cite[see][]{Shen et al.(2011)}.

\subsubsection{Comparison with previous works}
As shown in Fig. \ref{fig:allbald}, a strong anti-correlation has been found for the \mgii line when all the measurements from multi-epoch spectra are put together, suggesting a Baldwin effect for the whole sample. However, some scatters are also found, which is likely, at least partly, caused by the mixture of the intrinsic Baldwin effect and light-travel time effects in each object. The Baldwin effect in \mgii line is consistent with previous works. In the comprehensive study for DR7 quasar catalog (105,783 quasars), the strong Baldwin effects were found in both \mgii and \civ \citep{Shen et al.(2011)}. In contrast, the Baldwin effect in \hb is still unclear. The weak negative Baldwin effect was found for \hb in the forms of $ EW \propto L^{0.2}$, and $ EW \propto L^{0.1}$ from $\sim$ 22,000 quasars in 2dF+6dF surveys \citep{croom2002}, and $\sim$ 40,000 SDSS quasars \citep{greene2005,netzer2007}, respectively. It was suggested that this unexpected effect is likely caused by the luminosity-dependent change in the ratio of disc to non-disc continuum components \citep{croom2002}. Indeed, although our sample is rather small compared to their samples, we found a strong negative Baldwin effect in \hb $ EW \propto L^{0.14}$ with a Spearman correlation coefficient $r_{s} = 0.5$ at $\gg$ 99.9$\%$ confidence level, consistent with their results. The negative Baldwin effect between the broad $\rm H\beta$ EW and $L_{5100}$ has also been found below $L_{5100}\sim 10^{45} \rm erg~ s^{-1}$ in DR7 quasars \cite[see their Fig. 12,][]{Shen et al.(2011)}. However, the authors argued that it can be most likely due to the contamination from the host galaxy. When we restrict $L_{5100}\ge 10^{45} \rm erg~ s^{-1}$, we did not find pronounced Baldwin effect in $\rm H\beta$ for our sample. However, the small sample size with only 13 quasars at $L_{5100}\ge 10^{45} \rm erg~ s^{-1}$ precludes us to draw firm conclusions. Similarly, we failed to find Baldwin effect in \civ, which may be caused by the narrow $L_{1350}$ coverage \citep{Shields(2007)} (see Fig. 8).

\subsection{Line width}
An anti-correlation between the FWHM and luminosity of emission lines has been found in previous works \citep*[e.g.,][]{Wills et al.(1993)}. An intermediate-line region (ILR) is hypothesized to explain the anti-correlation, which is located between the narrow-line region and the very broad-line region, i.e., there are two distinct portions of the broad-line region. In the empirical relation of black hole mass estimations \cite[e.g.,][]{kaspi2000}, the black hole mass can be obtained from
\begin{equation}
M_{\rm BH} \propto V^2 \times R_{\rm BLR} \propto \rm FWHM^2 \times \it L^{\beta},
\end{equation}
where the $M_{\rm BH}$ is the black hole mass, V is velocity of the BLR gas clouds, $R_{\rm BLR}$ is BLR radius, $\beta$ is the index of the empirical $R_{\rm BLR}$ - $L$ relation, and $L$ is either the continuum or emission line luminosity \cite[e.g.,][]{wu04}. Therefore, an anti-correlation between emission line luminosity and the line width would be expected in individual sources, which indeed has been found in some quasars. However, we find positive correlations between FWHM and line luminosity in most sources (see Tables \ref{table:1} - \ref{table:3}). The positive correlations become even more evident when putting all quasars together in Fig. \ref{fig:allWHM}, which is consistent with \cite{Wilhite et al. (2005)}. This result can be explained with two components in BLR, with the broad line base (the inner BLR) more variable than the line core, which is expected because with the radius increasing, the characteristics of the variation will be diluted and the reprocessing of the seed photons will consume the energy of the variability \citep*[see][]{Wilhite et al. (2005)}. In the double-peaked broad \ha source 3C 390.3, \cite{zhang et al.(2013)} argued that this unexpected positive correlation can be naturally explained, due to the different time delays for different parts of the disk-like BLRs in one short period with the theoretical accretion disk model.

\section{Conclusions}
We have investigated the optical variability at various aspects for a sample of 60 SDSS quasars with at least six-epoch spectroscopic observations. The main results are summarized as follows:\\

(1) We verify the strong anti-correlation between the variability and continuum luminosity, consistent with previous works. A mild anti-correlation is discovered between variability and redshift. However, we find no significant correlations between the variability and black hole mass or Eddington ratio. In individual sources, half of the sample objects show BWB, while the rest half exhibit RWB. Although the mechanism of RWB is unclear, the effects of host galaxy contribution due to seeing variations can not be completely ruled out.

(2) As expected from photoionization model, the positive correlations between the broad emission line and continuum luminosity
are found in most individual sources, as well as for the whole sample. We confirm the Baldwin effect in most individual objects and the whole sample, while the negative Baldwin effect were also found in several quasars, which can be at least partly (if not all) due to the host galaxy contamination.

(3) We find positive correlations between the broad emission line luminosity and line width in most individual quasars, as well as the whole sample, implying a more variable line base than the line core.\\

We thank Zhaoyu Chen for help with data reduction. HXG thanks the Center for Astrophysics of USTC for teaching during his stay in USTC. This work is supported by the National Science Foundation of China (grant 11373056)
 and by the Science and Technology Commission of Shanghai Municipality (14ZR1447100). 

Funding for SDSS-III has been provided by the Alfred P. Sloan
Foundation, the Participating Institutions, the National Science
Foundation, and the U.S. Department of Energy Office of Science.
The SDSS-III web site is http://www.sdss3.org.

SDSS-III is managed by the Astrophysical Research Consortium for the
Participating Institutions of the SDSS-III Collaboration including the
University of Arizona,
the Brazilian Participation Group,
Brookhaven National Laboratory,
University of Cambridge,
Carnegie Mellon University,
University of Florida,
the French Participation Group,
the German Participation Group,
Harvard University,
the Instituto de Astrofisica de Canarias,
the Michigan State/Notre Dame/JINA Participation Group,
Johns Hopkins University,
Lawrence Berkeley National Laboratory,
Max Planck Institute for Astrophysics,
Max Planck Institute for Extraterrestrial Physics,
New Mexico State University,
New York University,
Ohio State University,
Pennsylvania State University,
University of Portsmouth,
Princeton University,
the Spanish Participation Group,
University of Tokyo,
University of Utah,
Vanderbilt University,
University of Virginia,
University of Washington,
and Yale University.


{}

\begin{landscape}
\begin{deluxetable}{c c c c c c c c c c c c c c c c c c c c c}
\setlength{\tabcolsep}{0.02in}
\tablewidth{0pt}
\tabletypesize{\footnotesize}

\tablecaption{Quasars at $z\leq$ 0.4
}
\tablehead{
\colhead{Object (SDSS J)}&
\colhead{$z$}&
\colhead{$i$}&
\colhead{N}&
\colhead{log $M_{\rm bh}$}&
\colhead{$\dot{m}$ }&
\colhead{$R$}&
\multicolumn{2}{c}{$L_{\rm con}$ - $\alpha_{\lambda}$}&
\multicolumn{2}{c}{$L_{\rm con}$ - $L_{\rm H\beta}$}&
\multicolumn{2}{c}{$L_{\rm con}$ - $L_{\rm H\alpha}$}&
\multicolumn{2}{c}{$L_{\rm con}$ - $EW_{\rm H\beta}$}&
\multicolumn{2}{c}{$L_{\rm con}$ - $EW_{\rm H\alpha}$}&
\multicolumn{2}{c}{$L_{\rm H\beta}$ - FWHM}&
\multicolumn{2}{c}{$L_{\rm H\alpha}$ - FWHM}\\
\cline{8-9}
\cline{10-11}
\cline{12-13}
\cline{14-15}
\cline{16-17}
\cline{18-19}
\cline{20-21}
\\[-2ex]
& &{(mag)}& &{$(M_{\odot})$}& & &{r}&{p}&{r}&{p}&{r}&{p}&{r}&{p}&{r}&{p}&{r}&{p}&{r}&{p}\\ [0.5ex]
\colhead{(1)}&\colhead{(2)}&\colhead{(3)}&\colhead{(4)}&\colhead{(5)}&\colhead{(6)}&\colhead{(7)}&\colhead{(8)}&\colhead{(9)}&
\colhead{(10)}&\colhead{(11)}&\colhead{(12)}&\colhead{(13)}&\colhead{(14)}&\colhead{(15)}&\colhead{(16)}&\colhead{(17)}&\colhead{(18)}
&\colhead{(19)}&\colhead{(20)}&\colhead{(21)}
}
\startdata
030639.57+000343.1&0.107&16.83&	9&	7.50&	0.24 	&5.3&	-0.89&	5E-4	&0.86&	1E-3&	0.86&	2E-3& 0.50 &0.17	&0.78	 &0.01	 & -0.32	 &0.35	 &-0.35	&0.31 \\[-0.1ex]
031027.82-004950.7& 0.081&15.63&	8	&7.92&	0.10&		-&	-0.59&	0.11&	0.92&	8E-4&	-0.14&	0.73&	-0.07&	0.86&	-0.78&	 0.02&	0.19&	 0.65&	 -0.66&	0.07  \\[-0.1ex]
031142.02-005918.9& 0.281&18.69&	6	&8.37&  0.04&	-	&0.37&	0.46&	0.60   &0.20   &     0.71	&0.11	&0.54&	0.26&	-0.48&	0.32&	 0.75&	 0.08&	 0.54&	 0.26  \\[-0.1ex]
031427.45-011152.3&0.387&18.33&	8&	7.86&	0.53&		-	&-0.16	&0.66&	0.73&	0.02&	0.63&	0.06&	-0.51	&0.15&	-0.46&	 0.21&	 0.13&	 0.73&	 0.66&	0.04  \\[-0.1ex]
\enddata
\label{table:1}
\tablecomments{The quasars at $z\leq$ 0.4 in our sample: Col. (1) object name; Col. (2) redshift; Col. (3) i-band apparent magnitude; Col. (4) number of spectroscopic observations; Col. (5) black hole mass; Col. (6) Eddington ratio, defined as $L_{\rm bol}/L_{\rm Edd}$; Col. (7) radio loudness, $R={f_{\rm 6 cm}/ f_{\rm 2500{\AA}}}$; Cols. (8) - (21) the Spearman rank correlation coefficients and probability level for $L_{\rm con}$ - $\alpha_{\lambda}$, $L_{\rm con}$ - $L_{\rm H\beta}$, $L_{\rm con}$ - $L_{\rm H\alpha}$, $L_{\rm con}$ - $EW_{\rm H\beta}$, $L_{\rm con}$ - $EW_{\rm H\alpha}$, $L_{\rm H\beta}$ - FWHM, and $L_{\rm H\alpha}$ - FWHM.}
\end{deluxetable}
\end{landscape}

\begin{landscape}

\begin{deluxetable}{c c c c c c c c c c c c c c c c c c c c c}

\setlength{\tabcolsep}{0.02in}
\tablewidth{0pt}
\tabletypesize{\footnotesize}
\tablecaption{Quasars at 0.4 $<  z \leq$ 0.8
}
\tablehead{
\colhead{Object (SDSS J)}&
\colhead{$z$}&
\colhead{$i$}&
\colhead{N}&
\colhead{log $M_{\rm bh}$}&
\colhead{$\dot{m}$ }&
\colhead{$R$}&
\multicolumn{2}{c}{$L_{\rm con}$ - $\alpha_{\lambda}$}&
\multicolumn{2}{c}{$L_{\rm con}$ - $L_{\rm MgII}$}&
\multicolumn{2}{c}{$L_{\rm con}$ - $L_{\rm H\beta}$}&
\multicolumn{2}{c}{$L_{\rm con}$ - $EW_{\rm MgII}$}&
\multicolumn{2}{c}{$L_{\rm con}$ - $EW_{\rm H\beta}$}&
\multicolumn{2}{c}{$L_{\rm MgII}$ - FWHM}&
\multicolumn{2}{c}{$L_{\rm H\beta}$ - FWHM}\\
\cline{8-21}
\\[-2ex]
& &{(mag)}& &{$(M_{\odot}$)}&&&{r}&{p}&{r}&{p}&{r}&{p}&{r}&{p}&{r}&{p}&{r}&{p}&{r}&{p}\\ [-0.5ex]
\colhead{(1)}&\colhead{(2)}&\colhead{(3)}&\colhead{(4)}&\colhead{(5)}&\colhead{(6)}&\colhead{(7)}&\colhead{(8)}&\colhead{(9)}&
\colhead{(10)}&\colhead{(11)}&\colhead{(12)}&\colhead{(13)}&\colhead{(14)}&\colhead{(15)}&\colhead{(16)}&\colhead{(17)}&\colhead{(18)}
&\colhead{(19)}&\colhead{(20)}&\colhead{(21)}
}
\startdata
021953.04-004434.2  &   0.686  &19.60& 6 &     8.25  & -0.76 & -  &   -0.08 & 0.87  & 0.43 &  0.39  & 0.20 &  0.70  & -0.54 & 0.27 &  -0.02 & 0.95 &  0.02 &  0.96 &  0.60 &  0.20 \\[-0.1ex]
022214.38-001745.3  &   0.773  &21.18& 6 &     8.42  & -1.06 & -  &   -0.60 & 0.28  & 0.89 &  0.03  & 0.80 &  0.10  & -0.70 & 0.19 &  0.50  & 0.39 & -0.20 &  0.74 &  0.40 &  0.50 \\[-0.1ex]
022331.90-001605.5  &   0.771  &19.00& 6 &     8.27  & -0.46 & -  &   0.09  & 0.87  & -0.10&  0.87  & 0.99 &  1E-4  & -0.90 & 0.03 &  -0.90 & 0.03 &  0.70 &  0.18 & -0.50 &  0.40 \\[-0.1ex]
022335.84+002351.8  &   0.774  &19.07& 9 &     8.25  & -0.64 & -  &   -0.11 & 0.76  &-0.08 &  0.83  & 0.25 &  0.52  & -0.08 & 0.83 &  -0.07 & 0.86 & -0.51 &  0.15 &  0.11 &  0.76 \\[-0.1ex]
022556.34+001345.3  &   0.709  &19.51& 7 &     8.81  & -1.12 & -  &   0.46  & 0.29  & 0.78 &  0.04  & 0.82 &  0.02  & -0.03 & 0.93 &  -0.11 & 0.82 &  0.82 &  0.02 &  0.57 &  0.18 \\[-0.1ex]
030458.96+000235.7  &   0.564  &18.36& 10&     9.03  & -1.17 &404.4&   -0.76 &9E-3  & 0.01 &  0.98  & 0.08 &  0.83  & -0.50 & 0.13 &  -0.43 & 0.21 &  0.83 &  3E-3 &  0.72 &  0.01 \\[-0.1ex]
030745.95+000833.4  &   0.427  &19.02& 9 &     8.50  &0.05   & -  &  -0.59  & 0.11  & -0.26&  0.53  & 0.26 &  0.50  & -0.45 & 0.26 &  -0.26 & 0.53 &  0.90 &  2E-3 &  0.59 &  0.11\\[-0.2ex]
030911.64+002358.8  &   0.611  &17.22& 10&     9.05  & -0.62 & -  &   -0.66 & 0.07  & 0.38 &  0.35  & 0.26 &  0.53  & -0.69 & 0.05 &  -0.21 & 0.61 &  0.97 &  4E-5 & -0.17 &  0.69 \\[-0.1ex]
030939.45-000339.2  &   0.769  &17.11& 8 &     8.92  & -0.36 & -  &   0.16  & 0.69  &-0.02 &  0.95  & 0.54 &  0.16  & -0.19 & 0.65 &  0.33  & 0.41 &  -0.88&  3E-3 &  0.73 &  0.03 \\[-0.1ex]
031022.10+004130.0  &   0.656  &19.47& 7 &     7.90  & -0.31 & -  &   0.67  & 0.09  & -0.46&  0.29  & 0.18 &  0.70  & -0.42 & 0.33 &  -0.32 & 0.48 &  0.96 &  1E-4 &  0.71 &  0.01 \\[-0.1ex]
031226.12-003708.9  &   0.621  &18.98& 10&     9.06  & -1.51 &1227.7& -0.06 & 0.86  & 0.28 &  0.46  & 0.92 & 5E-4 & -0.87 & 2E-3 &  -0.47 & 0.20   & -0.16 &  0.66 & -0.06 &  0.86 \\[-0.1ex]
032142.83-003225.7  &   0.648  &19.14& 6 &     8.17  & -0.46 & -  &   0.30  & 0.62  & 0.00 &  1.00   & 0.99 &  1E-4 & -0.70  & 0.18 &  -0.10 & 0.87 &  0.69 &  0.18 & -0.10 &  0.87 \\[-0.1ex]
032205.04+001201.4  &   0.471  &17.48& 6 &     8.70  & -0.71 & -  &   -0.71 & 0.11  & 0.42 &  0.39  & 0.94 &  4E-3 & -0.37  & 0.46 &   0.09 & 0.87 &  0.25 &  0.62 &  0.37 &  0.46 \\[-0.1ex]
\enddata
\label{table:2}
\tablecomments{The quasars at 0.4 $<  z \leq$ 0.8 in our sample: Col. (1) object name; Col. (2) redshift; Col. (3) i-band apparent magnitude; Col. (4) number of spectroscopic observations; Col. (5) black hole mass; Col. (6) Eddington ratio, defined as $L_{\rm bol}/L_{\rm Edd}$; Col. (7) radio loudness, $R={f_{\rm 6 cm}/ f_{\rm 2500{\AA}}}$; Cols. (8) - (21) the Spearman rank correlation coefficients and probability level for $L_{\rm con}$ - $\alpha_{\lambda}$, $L_{\rm con}$ - $L_{\rm MgII}$, $L_{\rm con}$ - $L_{\rm H\beta}$, $L_{\rm con}$ - $EW_{\rm MgII}$, $L_{\rm con}$ - $EW_{\rm H\beta}$, $L_{\rm MgII}$ - FWHM, and $L_{\rm H\beta}$ - FWHM.}
\end{deluxetable}
\end{landscape}

\begin{landscape}

\begin{deluxetable}{c c c c c c c c c c c c c c c c c c c c c}

\setlength{\tabcolsep}{0.02in}
\tablewidth{0pt}
\tabletypesize{\footnotesize}
\tablecaption{Quasars at $z>$ 0.8
}
\tablehead{
\colhead{Object (SDSS J)}&
\colhead{$z$}&
\colhead{$i$}&
\colhead{N}&
\colhead{log $M_{\rm bh}$}&
\colhead{$\dot{m}$ }&
\colhead{$R$}&
\multicolumn{2}{c}{$L_{\rm con}$ - $\alpha_{\lambda}$}&
\multicolumn{2}{c}{$L_{\rm con}$ - $L_{\rm C_{IV}}$}&
\multicolumn{2}{c}{$L_{\rm con}$ - $L_{\rm MgII}$}&
\multicolumn{2}{c}{$L_{\rm con}$ - $EW_{\rm C_{IV}}$}&
\multicolumn{2}{c}{$L_{\rm con}$ - $EW_{\rm MgII}$}&
\multicolumn{2}{c}{$L_{\rm C_{IV}}$ - FWHM}&
\multicolumn{2}{c}{$L_{\rm MgII}$ - FWHM}\\
\cline{8-21}\\[-2ex]
& &{(mag)}&&{$(M_{\odot})$}&&&{r}&{p}&{r}&{p}&{r}&{p}&{r}&{p}&{r}&{p}&{r}&{p}&{r}&{p}\\ [-0.5ex]
\colhead{(1)}&\colhead{(2)}&\colhead{(3)}&\colhead{(4)}&\colhead{(5)}&\colhead{(6)}&\colhead{(7)}&\colhead{(8)}&\colhead{(9)}&
\colhead{(10)}&\colhead{(11)}&\colhead{(12)}&\colhead{(13)}&\colhead{(14)}&\colhead{(15)}&\colhead{(16)}&\colhead{(17)}&\colhead{(18)}
&\colhead{(19)}&\colhead{(20)}&\colhead{(21)}
}
\startdata
021754.80+000234.0   &2.044 &19.07& 6  &    9.17 &0.22     &     -    & 0.37  & 0.46  & 0.82  & 4E-3  &-0.77 &  0.07 &  0.82 &  0.04 &  -0.88 & 0.02  & -0.65 &  0.15 & 0.60  & 0.20     \\[-0.2ex]
022111.89+010548.8   &1.487 &17.76& 7  &    9.35 &0.24     &     -    & 0.35  & 0.43  &  -    &  -    & 0.10  & 0.81 &   -   &   -   &  -0.57  & 0.18  &   -   &   -   & 0.39  & 0.38     \\[-0.2ex]
022143.19-001803.8   &2.621 &18.89& 7  &    9.64 &0.19     &     -    & 0.21  & 0.64  & 0.92  & 2E-3  &  -   &   -   &  0.42 &  0.33 &   -    &  -    & 0.92  & 2E-3  &  -    &  -       \\[-0.2ex]
022157.81+000042.5   &1.041 &18.39& 7  &    9.53 &0.04     &     -    &-0.07  & 0.87  &  -    &  -    & 0.89 &  6E-3 &   -   &   -   &  0.07  & 0.87  &   -   &   -   &-0.14  & 0.75     \\[-0.2ex]
022230.28+001844.5   &2.191 &19.04& 7  &    9.32 &0.15     &     -    &-0.39  & 0.38  &-0.61  & 0.14  & 0.00 &  1.00 & -0.53 &  0.21 &  0.11  & 0.82  & -0.85 & 0.01  & 0.64  & 0.12     \\[-0.2ex]
022246.46-004836.1   &1.541 &17.28& 7  &    9.10 &0.81     &     -    & 0.36  & 0.43  &  -    &  -    & 0.67 &  0.09 &   -   &   -   &  -0.35 & 0.43  &   -   &   -   &-0.03  & 0.93     \\[-0.2ex]
022321.38-000733.8   &1.534 &19.00& 6  &    8.96 &0.28     &     -    & 0.08  & 0.87  &  -    &  -    & 0.71 &  0.11 &   -   &   -   &  0.03 & 0.96  &   -   &   -   & 0.82  & 0.04     \\[-0.2ex]
022400.23-001241.3   &1.571 &20.01& 7  &    8.96 &0.10     &     -    &-0.82  & 0.04  &-0.02  & 0.95  & 0.60 &  0.21 & -0.42 &  0.39 &  -0.14 & 0.78  & 0.08  & 0.87  & 0.60  & 0.20     \\[-0.2ex]
022430.17-004131.1   &1.669 &18.94& 6  &    9.24 &0.13     &     -    & 0.25  & 0.62  &-0.25  & 0.62  &-0.14 &  0.78 & -0.31 &  0.54 &  -0.31 & 0.54  & 0.94  & 4E-3  & 0.77  & 0.07     \\[-0.2ex]
022518.36-001332.3   &3.628 &19.18& 11 &    9.36 &0.38     &     -    & 0.61  & 0.04  & 0.26  & 0.43  &  -   &   -   &-0.29  &  0.37 &    -   &  -    &-0.06  & 0.85  &  -    &  -       \\[-0.2ex]
022554.85+005451.9   &2.969 &18.65& 15 &    8.98 &1.10     &     -    & 0.09  & 0.73  & 0.33  & 0.22  &  -   &   -   & -0.33 &  0.21 &    -   &  -    & 0.33  & 0.22   &  -    &  -       \\[-0.2ex]
022826.69-003802.3   &1.515 &18.24& 8  &    9.48 &0.16     &     -    & 0.11  & 0.77  &  -    &  -    & 0.45 &  0.26 &   -   &   -   &  -0.12 & 0.78  &   -   &   -   & 0.19  & 0.65     \\[-0.2ex]
022844.09+000217.0   &2.706 &17.87& 15 &    9.46 &0.47     &     -    &-0.66  & 0.01  & 0.02  & 0.92  &  -   &   -   & -0.47 &  0.10 &    -   &  -    & 0.02  & 0.92  &  -    &  -       \\[-0.2ex]
025754.18+000506.4   &2.679 &19.17& 8  &    9.96 &0.07     &     89.7 & 0.26  & 0.53  & 0.35  & 0.38  &  -   &   -   &  0.23 &  0.57 &    -   &  -     & 0.00  & 1.00  &  -    &  -       \\[-0.2ex]
030551.14-000557.3   &0.839 &19.21& 10  &   8.36 &0.36     &     -    &-0.75  & 0.01  &  -    &  -    &-0.18 &  0.63 &   -   &   -   &  -0.67 & 0.04  &   -   &   -   & 0.00   & 1.00     \\[-0.2ex]
030719.91+004538.7   &1.903 &19.00& 9  &    8.93 &0.39     &     -    & 0.06  & 0.86  & 0.73  & 0.02  & 0.13 &  0.73 &  0.46 &  0.20 &  -0.70 & 0.03  & 0.67  & 0.04  & 0.48  & 0.18     \\[-0.1ex]
030815.84+010721.4   &1.217 &18.80& 9  &    8.45 &0.60     &     -    & 0.10  & 0.79  &  -    &  -    & 0.69 &  0.03 &   -   &   -   &  0.53  & 0.13  &   -   &   -   &0.48   & 0.18     \\[-0.2ex]
030905.37+005808.9   &1.518 &19.21& 7  &    9.11 &0.15     &     -    &-0.11  & 0.77  &  -    &  -    &-0.61 &  0.14 &   -   &   -   & -0.75  & 0.05  &   -   &   -   & 0.53  & 0.21     \\[-0.2ex]
030907.49+002419.0   &2.083 &18.88& 8  &    9.76 &0.08     &     -    & 0.66  & 0.07  & 0.42  & 0.28  & 0.42 &  0.28 &  0.28 &  0.49 &  0.35  & 0.36  & 0.40  & 0.32  & 0.14  & 0.74     \\[-0.2ex]
031003.01-004645.7   &2.115 &17.74& 9  &    9.19 &1.20     &     -    & 0.56  & 0.11  & 0.16  & 0.66  &0.51  &  0.15  & -0.45 &  0.22 &  0.26 & 0.48 & 0.31  & 0.41  & 0.96  & 2E-5  \\[-0.2ex]
031019.95+010111.5   &1.389 &17.56& 9  &    9.58 &0.22     &     -    &-0.20  & 0.60  &  -    &  -    & 0.85 & 3E-3 &   -   &   -   &  0.48  & 0.19  &   -   &   -   & 0.41  & 0.26     \\[-0.2ex]
031028.87-005326.2   &2.443 &18.43& 9  &    9.74 &0.22     &     -    &-0.25  & 0.51  & 0.19  & 0.61  &  -   &   -   &  0.01 &  0.96 &    -   &  -    &-0.40  & 0.28  &  -    &  -       \\[-0.2ex]
031030.90+000517.6   &1.229 &19.50& 9  &    8.51 &0.36     &     -    &-0.18  & 0.63  &  -    &  -    & 0.06 &  0.86 &   -   &   -   &  0.11 & 0.76  &   -   &   -   & 0.56  & 0.11     \\[-0.2ex]
031036.84+005521.7   &3.783 &19.50& 9  &    9.90 &0.15     &     -    & 0.40  & 0.28  & -0.01 & 0.96  &  -   &   -   & -0.13 &  0.73 &    -   &  -    &-0.01  & 0.96  &  -    &  -       \\[-0.2ex]
031037.63+004008.9   &1.273 &18.67& 8  &    9.04 &0.14     &     -    &-0.59  & 0.11  &  -    &  -    & 0.59 &  0.11 &   -   &   -   &  -0.35 & 0.38  &   -   &   -   & 0.02  & 0.95     \\[-0.2ex]
031118.52+002437.0   &0.863 &18.19& 8  &    8.71 &0.30     &     -    &-0.32  & 0.48  &  -    &  -    &-0.46 &  0.29 &   -   &   -   &  -0.75 & 0.05  &   -   &   -   &-0.07  & 0.87     \\[-0.2ex]
031127.55+005357.4   &1.764 &18.90& 9  &    9.38 &0.12     &     -    & 0.39  & 0.38  & 0.21  & 0.64  &-0.28 &  0.53 & -0.03 &  0.93 &  -0.46 & 0.29  & -0.07 & 0.88  & 0.67  & 0.09     \\[-0.2ex]
031129.29+005638.6   &1.507 &19.35& 8  &    9.29 &0.09     &     45.3 &-0.78  & 0.01  &  -    &  -    & 0.08 &  0.83 &   -   &   -   &  -0.43 & 0.24  &   -   &   -   & 0.24  & 0.51     \\[-0.2ex]
031131.41-002127.4   &1.572 &19.35& 6  &    9.71 &0.09     &     -    &-0.23  & 0.57  & 0.42  & 0.28  & 0.14 &  0.73 &  0.26 &  0.53 &  -0.52 & 0.18  & 0.95  &2E-4 & 0.40  & 0.31     \\[-0.2ex]
031156.45-004157.0   &0.955 &19.35& 9  &    9.15 &0.06     &     -    &-0.33  & 0.41  &  -    &  -    & 0.19 &  0.65 &   -   &   -   &  -0.92 &8E-4 &   -   &   -   & 0.09  & 0.82     \\[-0.2ex]
031227.13-003446.2   &1.776 &18.88& 10 &    9.52 &0.08     &     -    &-0.28  & 0.42  & 0.34  & 0.32  & 0.41 &  0.37 &  0.16 &  0.65 &  0.46  & 0.17  & 0.26  & 0.47  & 0.72  & 0.01     \\[-0.2ex]
031237.56+004511.3   &1.822 &19.03& 9  &    9.49 &0.09     &     -    &-0.16  & 0.66  & 0.64  & 0.05  & 0.38 &  0.30 & -0.31 &  0.40 &  -0.37   &0.34    & -0.15 & 0.70  & 0.36  & 0.33     \\[-0.2ex]
031246.45-005024.6   &1.581 &18.75& 9  &    9.16 &0.22     &     -    & 0.40  & 0.28  &0.19   & 0.61  & 0.53 &  0.13 &  0.24 &  0.51 &  0.33  & 0.38  & 0.48  & 0.18  & 0.39  & 0.28     \\[-0.2ex]
031307.92-003221.9   &1.411 &19.04& 8  &    8.89 &0.23     &     -    &-0.13  & 0.75  &  -    &  -    &-0.19 &  0.64 &   -   &   -   &  -0.53  & 0.17  &   -   &   -   & 0.90  &2E-3    \\[-0.2ex]
031318.66+003623.9   &1.256 &18.07& 10  &   10.02&0.03     &     221.9&-0.68  & 0.02  &  -    &  -    &-0.45 &  0.18 &   -   &   -   & -0.45  & 0.18  &   -   &   -   & 0.97  & 1E-6     \\[-0.2ex]
031343.07-001623.4   &1.562 &19.32& 10  &   9.19 &0.12     &     -    &-0.09  & 0.80  & 0.50  & 0.13  & 0.91 &  2E-4 &  0.23 &  0.51 &  0.49  & 0.45  & 0.53  & 0.11  &-0.04  & 0.90     \\[-0.2ex]
031348.34-010433.0   &2.468 &18.99& 9  &    9.03 &0.60     &     -    & 0.35  & 0.35  & 0.54  & 0.12  &  -   &   -   &  0.44 &  0.22 &    -   &  -    & 0.54  & 0.12  &  -    &  -       \\[-0.2ex]
031404.44-003947.3   &2.105 &18.51& 9  &    9.80 &0.12     &     -    & 0.46  & 0.20  & 0.34  & 0.35  & 0.78 &  0.01 & 0.23 &  0.54 &  0.03 & 0.92  & 0.00  & 1.00  & 0.10  & 0.79     \\[-0.2ex]
031439.08-000249.3   &1.152 &18.77& 8  &    8.65 &0.28     &     -    & 0.59  & 0.11  &  -    &  -    & 0.50 &  0.20 &   -   &   -   &  -0.67 & 0.07 &   -   &   -   & 0.80  & 0.01     \\[-0.2ex]
031444.54-005701.1   &1.506 &18.33& 9  &    9.43 &0.14     &     -    & 0.05  & 0.89  &  -    &  -    & 0.31 &  0.40 &   -   &   -   &  -0.26 & 0.48  &   -   &   -   & 0.14  & 0.70     \\[-0.2ex]
031452.06+001346.3   &3.202 &19.27&8  &    8.77 &1.73     &     66.4 &-0.65  & 0.15  & 0.14  & 0.78  &  -   &   -   & -0.25 &  0.62 &    -   &  -    & 0.14  & 0.78  &  -    &  -       \\[-0.2ex]
031645.55-000553.3   &1.397 &19.33& 8  &    9.72 &0.03     &     18.4 &-0.6   & 0.14  &  -    &  -    & 0.00 &  1.00 &   -   &   -   &  -0.28 & 0.53  &   -   &   -   & 0.73  & 0.06     \\[-0.2ex]
032933.97-004801.0   &1.878 &18.83& 6  &    9.07 &0.46     &     -    &-0.77  & 0.07  & 0.88  & 0.02  & 0.48 &  0.32 &  0.60 &  0.21 &   -0.02 &0.96   & -0.43 &  0.39 & 0.94  & 4E-3    \\[-0.2ex]
\enddata
\label{table:3}
\tablecomments{The quasars at $z>$ 0.8 in our sample: Col. (1) object name; Col. (2) redshift; Col. (3) i-band apparent magnitude; Col. (4) number of spectroscopic observations; Col. (5) black hole mass; Col. (6) Eddington ratio, defined as $L_{\rm bol}/L_{\rm Edd}$; Col. (7) radio loudness, $R={f_{\rm 6 cm}/ f_{\rm 2500{\AA}}}$; Cols. (8) - (21) the Spearman rank correlation coefficients and probability level for $L_{\rm con}$ - $\alpha_{\lambda}$, $L_{\rm con}$ - $L_{\rm C_{\rm IV}}$, $L_{\rm con}$ - $L_{\rm MgII}$, $L_{\rm con}$ - $EW_{\rm C_{\rm IV}}$, $L_{\rm con}$ - $EW_{\rm MgII}$, $L_{\rm C_{\rm IV}}$ - FWHM, and $L_{\rm MgII}$ - FWHM.}
\end{deluxetable}
\end{landscape}

\end{CJK}
\end{document}